%% file: main.tex
\documentclass[lettersize,journal]{IEEEtran}
\usepackage{graphicx}
\usepackage{tikz}
\usepackage{pgfplots}
\usepackage{psfrag}
\usepackage[justification=centering]{caption}
\usetikzlibrary{arrows,snakes,shapes,shapes.geometric,fit,positioning,arrows,decorations.markings,calc}
\tikzstyle{block} = [rectangle, draw, fill=blue!20, 
text width=5em, text centered, rounded corners, minimum height=4em]
\tikzstyle{line} = [draw, -latex']
\usepackage{mathtools}
\usepackage{amsbsy}
\usepackage{amsmath}
\usepackage{fixmath}
\usepackage{amssymb}
\usepackage{mathrsfs}
\usepackage{fancyhdr}
\usepackage{xcolor}
\usepackage{hyperref}
\usepackage{algorithm}
\usepackage{algcompatible}
\usepackage{algpseudocode}
\usepackage{multirow}
\newcommand{\bs}[1]{\boldsymbol{#1}}
\newcommand{\tc}{\text{c}}
\newcommand{\tp}{\text{p}}
\newcommand{\HH}{\mathrm{H}}
\newcommand{\TT}{\mathrm{T}}

\usepackage{varwidth}

\definecolor{lime}{HTML}{A6CE39}
\DeclareRobustCommand{\orcidicon}{%
	\begin{tikzpicture}
	\draw[lime, fill=lime] (0,0) 
	circle [radius=0.16] 
	node[white] {{\fontfamily{qag}\selectfont \tiny ID}};
	\draw[white, fill=white] (-0.0625,0.095) 
	circle [radius=0.007];
	\end{tikzpicture}
	\hspace{-2mm}
}

\foreach \x in {A, ..., Z}{%
	\expandafter\xdef\csname orcid\x\endcsname{\noexpand\href{https://orcid.org/\csname orcidauthor\x\endcsname}{\noexpand\orcidicon}}
}

\ifCLASSINFOpdf
\else
\fi

\newcommand{\copyrighttext}{%
\footnotesize\textcopyright This work has been submitted to the IEEE for possible publication. Copyright may be transferred without notice, after which this version may no longer be accessible.}

\newcommand{\copyrightnotice}{%
\begin{tikzpicture}[remember picture,overlay]
\node[anchor=south,yshift=10pt] at (current page.south) {\fbox{\parbox[]{\dimexpr\textwidth-\fboxsep-\fboxrule\relax}{\copyrighttext}}};
\end{tikzpicture}%
}

\begin{document}

\title{Rate Splitting in FDD Massive MIMO Systems Based on the Second Order Statistics\\ of Transmission Channels}

\author{Donia~Ben~Amor\orcidA{},~\IEEEmembership{Student~Member,~IEEE,} \and
        Michael~Joham\orcidB{},~\IEEEmembership{Member,~IEEE,}\and \\
       and Wolfgang~Utschick\orcidC{},~\IEEEmembership{Fellow,~IEEE}
        \thanks{This work was supported by the German Research Foundation under Grant UT 36/18-1}
}

% use section* for acknowledgment

% make the title area
\maketitle
\vspace*{-0.75cm}

\copyrightnotice

% As a general rule, do not put math, special symbols or citations
% in the abstract or keywords.
\begin{abstract}
In this work, we present new results for the application of rate splitting multiple access (RSMA) to the downlink (DL) of a massive multiple-input-multiple-output system (MIMO) operating in frequency-division-duplex (FDD) mode. We propose a statistical precoding design relying on the channels' second-order information when one-layer RS is implemented. The advantage of the statistical precoding lies in simplifying the precoder design by basing the optimization objective on the slowly-varying channel covariance matrices instead of the fast-changing instantaneous channel estimates.
To this end, we use the so-called bilinear precoder, which was shown in \cite{Bilinear} and \cite{Donia_WSA20} to have limited performance in the high SNR regime due to the imperfect channel state information (CSI) available at the base station (BS). We formulate the DL throughput maximization problem based on a widely used lower bound on the achievable sum rate and propose an iterative approach to solve the underlying optimization problem. Numerical results demonstrate the benefit of implementing one-layer RS. Furthermore, the proposed iterative approach achieves excellent results in terms of spectral efficiency compared to the state-of-the-art techniques.

\end{abstract}

% Note that keywords are not normally used for peerreview papers.

\section{Introduction}
Massive multiple-input-multiple-output (MaMIMO) systems are a prominently used technology in current wireless communication standards due to the several advantages ranging from increased spectral efficiency to improved energy efficiency and reliability to name a few \cite{mimo}, \cite{marzetta}. These benefits are closely related to the presence of an accurate channel state information (CSI) at the base station (BS). The acquisition of the CSI is, however, often a challenging task due to the typically very small channel coherence interval \cite{Adhikary}. An imperfect CSI can tremendously impact the system throughput especially in the high SNR regime where the system becomes interference limited due to the CSI systematic estimation errors. While the channels' second-order statistics could be exploited to mitigate the interference, the latter cannot be completely removed and it has been observed in several works on statistical precoding/equalization, e.g., \cite{Bilinear} and \cite{Donia_WSA20} that the achievable rate saturates in the high SNR regime due to pilot contamination. The latter effect results from pilot reuse among users during channel probing of a time-division-duplex (TDD) system. A similar effect is observed in FDD systems, when the number of orthogonal pilots used for training the DL channel is less than the number of BS antennas. In either cases, the channel estimates suffer from a systematic error, even when investing high transmit power during training \cite{massiveMIMObook}.
\par In this work, we focus on a MaMIMO system operating in FDD mode, where the CSI acquisition is more challenging than in TDD systems. This mainly results from the lack of channel reciprocity between the uplink (UL) and DL channels due to UL-DL frequency gap being larger than the channel coherence bandwidth. Hence, explicit training of the DL channel and subsequent feedback are inevitable. On the one hand, the length of the training period should be at least equal to the number of BS antennas to ensure a contamination-free estimation, which on the other hand results in a huge training overhead considering the assumed massive number of BS antennas $M\gg 1$. We, therefore, shrink the length of the training phase to $T_\text{dl}<M$. This obviously comes at the cost of a contaminated channel observation at the user side, i.e., interference between the channel coefficients. The goal is to design a transmit strategy at the BS that takes into account the underlying systematic error in the channel estimates.
\par One of the promising approaches to achieve higher data rates under imperfect CSI is RSMA. This technique combines the advantages of conventional broadcasting and multicasting systems, where the transmitter (BS) splits the message intended to each user into common and private parts. Each user has then to decode the common messages first and then apply successive interference cancellation to remove the interference from the common part. Depending on how many common streams are sent, an SIC layer per common stream is required at the receiver's side. In order to reduce the user's hardware complexity, we assume that one-layer RS is implemented, such that only one common stream is transmitted to all users. This implies that only one SIC layer is required at the user's side. In \cite{Dai}, an RS approach is proposed for a system with imperfect CSI at the transmitter where an adaptive power allocation strategy is developed. Furthermore, the authors suggest a hierarchical RS method where the long term CSI (channel covariance matrices) and the short-term CSI (instantaneous imperfect channel estimate) are exploited to design a two-stage precoder. The work \cite{Dizdar} considers a setup with user mobility. A lower bound on the ergodic sum rate is derived for this setup and a power allocation strategy is proposed for the 1-layer RS based system. In \cite{Mishra}, the authors show the ability of RS to mitigate the pilot contamination effect resulting from pilot reuse among users during the channel probing of a TDD massive MIMO system. Another application for RS in the case of imperfect CSI was investigated in \cite{Papazafeiropoulos}, where RS is used to alleviate the effect of hardware impairments on MaMIMO systems. In \cite{Clerckx_RS2} and \cite{IWMMSE}, a MaMIMO system with imperfect CSI at the transmitter is considered. The authors propose to use the so-called iterative weighted minimum mean squared error (IWMMSE) approach to solve the sum rate maximization problem for a linearly precoded DL system featuring one-layer RS. Although this approach has been shown to achieve very good performance under imperfect CSI at the transmitter, it remains computationally expensive, considering that one has to solve the underlying optimization problem in each channel coherence interval to find the precoders for the given instantaneous channel estimate. 
\par \textit{Contributions:}
We propose an alternative approach to the IWMMSE method in order to solve the throughput maximization problem, where we formulate the optimization problem depending on the channel statistics by combining the one-layer RS strategy with the bilinear precoding approach proposed in \cite{Donia_WSA20}. This precoder consists of a linear transformation of the received training signal, where the transformation is a deterministic matrix that is designed only based on the channel distribution information (CDI). Due to the slower variation of the channel statistics compared to the channel itself, one has hence to solve the optimization problem for the deterministic matrices only when the covariance matrices change, which tremendously reduces the computational complexity related to the precoder design. To this end, it is assumed that the channel covariance matrices or at least estimates thereof are available at the BS. We refer the reader to the works \cite{Extrapolation2}, \cite{Donia_WSA20}, where an extrapolation approach is used to acquire the DL channel covariance matrices from the UL channel covariance matrices by assuming the reciprocity of the so-called angular scattering function. As for the UL channel covariance matrices, those can be estimated using the UL training samples that are sent anyway for the purpose of UL channel estimation. In Section~\ref{sec5}, we study through numerical simulations the impact of imperfect channel covariance matrices on the system performance.\\
The formulated DL throughput maximization problem is involved and thus we opt for a decomposition of the underlying problem into three parts that are solved consecutively. We propose efficient solution approaches for each subproblem and use results on fractional programming proposed in \cite{FracProg}, \cite{WSRM} to make the optimization more tractable.
Despite the suboptimality of the proposed approach, it still leads to a satisfactory performance in terms of the achievable data rates, as the simulation results show in a later part of this work.
\par \textit{Paper Organization:} This paper is organized as follows. In Section~ \ref{sec1}, we present the preliminaries and the system setup considered in this work. In Section~ \ref{sec2}, we derive expressions for the considered utility function and the different constraints and formulate the optimization problem. Section~\ref{sec3} deals with the solution approach for the underlying optimization problem. In Section~\ref{sec4}, we present the performance of the proposed approach in terms of the spectral efficiency. A conclusion for this work is drawn in Section~\ref{sec5}. 
\par \textit{Notation:} Bold letters indicate vectors and matrices, non-bold
letters express scalars. The operators $(.)^{\star}$, $(.)^\TT$ and $(.)^\HH$
stand for complex conjugation, transposition and Hermitian
transposition, respectively. The square root matrix of a matrix $\bs{A}$ is denoted by $\bs{A}^\frac{1}{2}$ and satisfies $\bs{A}=\bs{A}^{\frac{1}{2},\HH}\bs{A}^\frac{1}{2}$. The $n\times n$ identity matrix is denoted by $\bs{I}_n$ and the vector $\bs{e}_m$ represents a zero-vector with 1 at the $m$th position. \\$\text{tr}(.)$ and $\otimes$ denote respectively the trace operator and the kronecker product, whereas $\text{vec}(.)$ performs a vectorisation of the input matrix by stacking its columns into a vector. 
$\text{blkdiag}(\bs{A},\bs{B})$ creates a block diagonal out of the matrices $\bs{A}$ and $\bs{B}$. Finally, we denote by $\text{var}$ and $\mathbb{E}$ the variance and the mean of a random variable, respectively.

\section{System Model}
\label{sec1}
We consider the DL of a single-cell massive multi-user multiple-input-single-output (MISO) system operating in FDD mode, where the BS is equipped with $M\gg 1$ antennas and serves $K$ single-antennas users. We denote by $\bs{h}_k\sim \mathcal{N}_\mathbb{C}\left(\bs{0},\bs{C}_k \right) \in \mathbb{C}^M$ the DL channel between the BS and user $k$, where $\bs{C}_k$ is the corresponding channel covariance matrix. We assume a block-fading channel model such that the channels remain unchanged for an interval of $T_\text{ch}$ time-frequency symbols, also known as the channel coherence interval.

The DL operation consists of a training phase and a data transmission phase. During the former, the BS station sends $T_\text{dl}<M$ orthogonal pilot sequences to all users, so that the $T_\text{dl}$ training symbols collected at user $k$ can be stacked in a column vector, which constitutes the observation vector $\bs{y}_k$ 
\begin{equation}
    \bs{y}_k=\bs{\Phi}^\HH\bs{h}_k+\bs{n}_k.
    \label{eq:yk}
\end{equation}
Here, $\bs{\Phi}\in \mathbb{C}^{M\times T_\text{dl}}$ denotes the pilot matrix, whose columns are given by the $T_\text{dl}$ orthogonal pilot sequences. This implies that $\bs{\Phi}^\HH\bs{\Phi}=\bs{I}_{T_\text{dl}}$, but $\bs{\Phi}\bs{\Phi}^\HH\neq\bs{I}_{M}$. The noise vector $\bs{n}_k\sim \mathcal{N}_\mathbb{C}\left(\bs{0},\sigma^2_n\bs{I}_{T_\text{dl}}\right)$ comprises the portion of noise resulting from the noisy DL training. Following the model in \cite{Extrapolation2}, we assume that the observation vector $\bs{y}_k$ is fed back to the BS in the UL through an analog feedback channel \cite{AnalogFB}, such that $\bs{n}_k$ also contains the noise portion modeling the feedback errors.

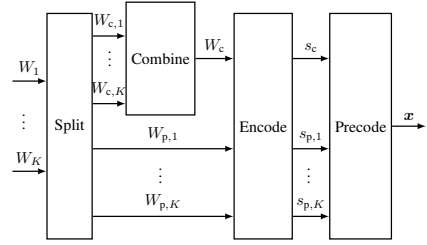
\begin{figure}
    \centering
    \scalebox{0.6}{\input{SystemModelTx}}
    \caption{System model at the transmitter side during the data transmission phase for the one-layer RS setup}
    \label{fig:SystemModelTx}
\end{figure}
\par Considering the DL data transmission phase, we assume that one-layer RS is implemented, i.e., only one layer of common messages for the $K$ users is used, [see \cite{RSMA}, Fig.1].\\
As depicted in Fig.~\ref{fig:SystemModelTx}, the message $W_k$ intended for some user $k$ is divided into a common part $W_{\tc,k}$ and a private part $W_{\tp,k}$. The common parts of all users' messages are combined to form a super common message $W_\tc$. The resulting $K+1$ messages, i.e., the super common message $W_\tc$ and the $K$ private messages $W_{\tp,1}, \dots, W_{\tp,K}$, are encoded independently into the symbols $s_{\tc}$, $s_{\tp,1} ,\dots, s_{\tp,K}$.\\
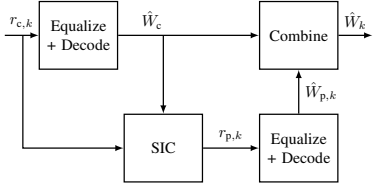
\begin{figure}[h]
    \centering
    \scalebox{0.6}{\input{SystemModelMS}}
    \caption{System model at some generic user $k$ during the data transmission phase for the one-layer RS setup}
    \label{fig:SystemModelMS}
\end{figure}
At the receiver side, we assume that SIC is implemented. Each user decodes firstly the common message $\hat{W}_\tc$, then subtracts the common message from the received signal through SIC and decodes its own private message $\hat{W}_{\tp,k}$ afterwards. The estimated message $\hat{W}_k$ is constructed by combining $\hat{W}_\tc$ and $\hat{W}_{\tp,k}$ (cf. Fig~\ref{fig:SystemModelMS}). \\
In order for the users to perfectly decode the common message, the common rate $R_{\tc}$ should not exceed the minimum common rate among all users, i.e., $R_{\tc}=\underset{k}{\min} \: R_{\tc,k}$.

Denoting by $\bs{x}$ the precoded transmit vector, the received signal at user $k$ is given by
\begin{equation}
    r_k=\bs{h}_k^\HH \bs{x} + v_k 
    \label{eq:rk}
\end{equation}
where $v_k\sim\mathcal{N}_\mathbb{C}(0,1)$ is the zero-mean unit-variance additive white Gaussian noise at the user side. 

Given the common precoder $\bs{p}_{\tc}$ and the private precoders $\bs{p}_{\tp,i}, i=1\dots K$, the transmit vector $\bs{x}$ can be written as
\begin{equation}
    \bs{x}=\bs{p}_{\tc} s_{\tc} +\sum_{i=1}^{K} \bs{p}_{\tp,i} s_{\tp,i}
	\label{eq:pci}
\end{equation}
where $s_{\tc}\sim \mathcal{N}_\mathbb{C}(0,1)$ and $s_{\tp,i}\sim \mathcal{N}_\mathbb{C}(0,1), i=1,\dots,K$.

Our aim is to design the common precoder $\bs{p}_{\tc}$ and the private precoder $\bs{p}_{\tp,k}$ for some user $k$ in the following form
\begin{align}
    \bs{p}_{\tc}&=\bs{A}_{\tc} \bs{y} \label{eq:pc} \\
    \bs{p}_{\tp,k}&=\bs{A}_{\tp,k} \bs{y}_k. \label{eq:pk}
\end{align}
In \eqref{eq:pc}, we introduced the vector $\bs{y}$ of all users' channel observations, i.e.,
\begin{equation}
    \bs{y}=[\bs{y}_1^\TT,\bs{y}_2^\TT,\dots, \bs{y}_K^\TT]^\TT
\end{equation}
where the $\bs{y}_k$ are defined in \eqref{eq:yk}. $\bs{A}_{\tc}$ and $\bs{A}_{\tp,k}$ in \eqref{eq:pc} and \eqref{eq:pk}, respectively, are deterministic transformation matrices which depend solely on the second-order channel statistics and on the training matrix $\bs{\Phi}$. These matrices are our design variables throughout this work.\\
The structure of the common precoder in \eqref{eq:pc} implies that the latter has to be designed as the superposition of "individual precoders", i.e., $\bs{p}_{\tc}=\sum_{k=1}^K \bs{A}_{\tc,k} \bs{y}_k$ with $\bs{A}_\tc=[\bs{A}_{\tc,1},\bs{A}_{\tc,2},\dots, \bs{A}_{\tc,K}]$. This choice is motivated by the multicast channel discussion in \cite{Hunger}. 

We refer to the precoders defined in \eqref{eq:pc} and \eqref{eq:pk} as bilinear precoders. We use this term because the received signal $r_k$ at user $k$ [cf. \eqref{eq:rk}] is not only linear in the data signals  $s_\tc$ and $s_{\tp,k}$, but also in the channel observation vector $\bs{y}_k$ [cf. \eqref{eq:yk}] due to \eqref{eq:pc} and \eqref{eq:pk}.

Due to the limited power budget available at the BS denoted by $P_\text{dl}$, the transmit vector $\bs{x}$ must satisfy the average transmit power constraint
\begin{equation}
    \mathbb{E}[\|\bs{x}\|^2]=\|\bs{p}_{\tc}\|^2 +\sum_{i=1}^{K}\|\bs{p}_{\tp,i}\|^2 \leq P_\text{dl}.
    \label{eq:powconst}
\end{equation}
Now, let $\alpha_{\tc}$ denote the power fraction allocated to the common stream transmission. Hence, the power constraint in \eqref{eq:powconst} can be reformulated as the combination of the following two constraints
\begin{equation}
\begin{aligned}
     \|\bs{p}_{\tc}\|^2&\leq \alpha_\tc P_\text{dl}\\ \sum_{i=1}^{K}\|\bs{p}_{\tp,i}\|^2 &\leq (1-\alpha_\tc) P_\text{dl}.
\end{aligned}
\label{eq:powconst2}
\end{equation}
As we aim at optimizing the precoders based on the second-order statistics of the channel, we relax the power constraints by taking the expectation with respect to the channels over the expressions in \eqref{eq:powconst2}. By inserting the expressions of the precoders \eqref{eq:pc} and \eqref{eq:pk} into the relaxed power constraints, we obtain
\begin{align}
     \mathbb{E}[\|\bs{p}_{\tc}\|^2]&=\mathrm{tr}\left(\bs{A}_{\tc}\bs{C}_{\bs{y}}\bs{A}_{\tc}^\HH\right) \leq \alpha_\tc P_\text{dl} \label{eq:ComPowCons}\\
     \sum_{i=1}^{K}\mathbb{E}[\|\bs{p}_{\tp,i}\|^2] &=\sum_{i=1}^{K}\mathrm{tr}\left( \bs{A}_{\tp,i} \bs{C}_{\bs{y}_i}\bs{A}_{\tp,i}^\HH\right) \leq (1-\alpha_\tc) P_\text{dl}\label{eq:PriPowCons}
\end{align} 
where we introduced $\bs{C}_{\bs{y}}$ in \eqref{eq:ComPowCons}, the covariance matrix of $\bs{y}$ given by
\begin{equation}
    \bs{C}_{\bs{y}}=\text{blkdiag}\left(\bs{C}_{\bs{y}_1}, \bs{C}_{\bs{y}_2},\dots, \bs{C}_{\bs{y}_K}\right)
\end{equation}
and $\bs{C}_{\bs{y}_k}$ is defined as the covariance matrix of $\bs{y}_k$, i.e.,
\begin{equation}
\bs{C}_{\bs{y}_k}= \bs{\Phi}^\HH\bs{C}_k\bs{\Phi}+\sigma_n^2\bs{I}_{T_\text{dl}}.
\label{eq:Cyk}
\end{equation} 
The block-diagonal structure of $\bs{C}_{\bs{y}}$ follows from the independence of the users' channel observations.

\section{Problem Formulation}
\label{sec2}
Throughout this work, our goal is to optimize the system throughput with respect to the deterministic transformations $\bs{A}_\tc$ and $\bs{A}_{\tp,k}, \forall k$, introduced in \eqref{eq:pc} and \eqref{eq:pk}, respectively. Note that since these transformations only depend on the channel covariance matrices, the considered optimization problem has to be only run, whenever the channel covariance matrices change, which happens much less frequently than the channel itself changes. Hence, the computation of each precoder consists simply of a matrix-vector multiplication in each channel coherence interval [cf. \eqref{eq:pc}, \eqref{eq:pk}]. 
\par We start this section by specifying the design objective. To this end, we derive closed-form expressions, which will allow us to formulate the optimization problem to be solved in Section \ref{sec3}.
\par Considering the RS setup, we deal with two kinds of messages: common and private messages. We, therefore, differentiate between the common rate $R_{\tc,k}$ and the private rate $R_{\tp,k}$ for each user $k$. 
\begin{equation}
\begin{aligned}
    &R_{\tc,k}= I(r_{\tc,k};s_\tc), \quad
    &R_{\tp,k}= I(r_{\tp,k};s_{\tp,k})
\end{aligned}
\label{eq:RcRp}
\end{equation}
where $I(r;s)$ denotes the mutual information between the two signals $r$ and $s$.\\
In \eqref{eq:RcRp}, we introduced $r_{\tc,k}$, the received signal for the common message at user $k$, which is given by
\begin{equation}
\begin{aligned}
     r_{\tc,k}&=\bs{h}_k^\HH \bs{p}_\tc s_\tc + \sum_{i=1}^K
    \bs{h}_k^\HH \bs{p}_{\tp,i} s_{\tp,i} + v_k\\
     &= b_{\tc,k} s_\tc + \sum_{i=1}^K b_{\tp,k,i} s_{\tp,i} +v_k.
\end{aligned}
\label{eq:rck}
\end{equation}
Here, the scalars $b_{\tc,k}=\bs{h}_k^\HH \bs{p}_\tc$ and $b_{\tp,k,i}=\bs{h}_k^\HH \bs{p}_{\tp,i}$ are the effective channels when incorporating the common and the private precoders, respectively.\\
In this work, we assume that user $k$ knows perfectly the effective channel coefficient $b_{\tc,k}$. This can be achieved by sending beamformed dedicated pilot symbols in the DL. This means, in addition to the matrix $\bs{\Phi}$, the BS sends precoded  pilots to the users, which can then estimate the combination of the channel with precoder given by the scalars $b_{\tc,k}$ and $b_{\tp,k,i}$. The knowledge of $b_{\tc,k}$ at the user's side is necessary to be able to perform perfect SIC, i.e., in order to decode the common message and subtract it entirely from the received signal.
After performing perfect SIC, we can write the received signal for the private message $r_{\tp,k}$ as follows
\begin{equation}
\begin{aligned}
     r_{\tp,k}&= \bs{h}_k^\HH \bs{p}_{\tp,k} s_{\tp,k}  +\sum_{\substack{i=1,\\ i\neq k}}^K
    \bs{h}_k^\HH \bs{p}_{\tp,i} s_{\tp,i} + v_k\\
    &=b_{\tp,k,k} s_{\tp,k} +\sum_{\substack{i=1,\\i\neq k}}^K b_{\tp,k,i} s_{\tp,i} +v_k.
\end{aligned}
\label{eq:rpk}
\end{equation}
\par Due to the imperfect channel state information at the transmitter side [cf. \eqref{eq:yk}], no closed-form expressions can be found for the achievable rates of the $R_{\tc,k}$ and $R_{\tp,k}$.
 
For such a setup, it is very common to consider a lower bound on the mutual information as the design objective. One widely used lower bound in the MaMIMO literature is the so-called hardening bound, which first appeared in \cite{Medard} and was then also investigated in \cite{marzetta2016fundamentals}.\\
For this lower bound, we assume that the BS only knows the statistics of the effective channels $b_{\tc,k}$ and $b_{\tp,k,i}$ [cf. \eqref{eq:rck}]. The estimates of these effective channels are simply given by their means
\begin{equation}
    \hat{b}_{\tc,k}=\mathbb{E}[b_{\tc,k}], \quad \hat{b}_{\tp,k,i}=\mathbb{E}[b_{\tp,k,i}].
\end{equation}
The effective channels can thus be written as
\begin{equation}
    b_{\tc,k}=\hat{b}_{\tc,k}+\Tilde{b}_{\tc,k}, \quad b_{\tp,k,i}=\hat{b}_{\tp,k,i}+\Tilde{b}_{\tp,k,i}
    \label{eq:bckbpk}
\end{equation}
with the zero-mean random parts $\Tilde{b}_{\tc,k}$ and $\Tilde{b}_{\tp,k,i}$.\\
Inserting the definitions of $b_{\tc,k}$ and $b_{\tp,k,i}$ into \eqref{eq:rck} and \eqref{eq:rpk}, we obtain the two portions of the received signal 
\begin{equation}
    \begin{aligned}
    r_{\tc,k}&=\hat{b}_{\tc,k} s_\tc + \underbrace{\Tilde{b}_{\tc,k} s_\tc + \sum_{i=1}^K b_{\tp,k,i} s_{\tp,i} + v_k}_{\eta_{\tc,k}}\\
    r_{\tp,k}&=\hat{b}_{\tp,k,k} s_{\tp,k} + \underbrace{\Tilde{b}_{\tp,k,k} s_{\tp,k} + \sum_{\substack{i=1\\ i\neq k}}^K b_{\tp,k,i} s_{\tp,i} + v_k}_{\eta_{\tp,k}}.
    \end{aligned}
    \label{eq:eta}
\end{equation}
\par The lower bound is essentially based on treating the effective noise terms $\eta_{\tc,k}$ and $\eta_{\tp,k}$ constituted by the interference and noise as a Gaussian uncorrelated additive noise, which corresponds to the worst-case noise. In several previous works (e.g., \cite{ieeCaire17}), such a lower bound for MaMIMO systems has been rigorously derived and with similar steps, the lower bounds for the common and private rates are given by
\begin{equation}
    \begin{aligned}
    I(r_{\tc,k};s_\tc)&\geq R_{\tc,k}^\text{LB}= \log_2(1+\gamma_{\tc,k}) \\
    I(r_{\tp,k};s_{\tp,k})&\geq R_{\tp,k}^\text{LB}= \log_2(1+\gamma_{\tp,k})
    \end{aligned}
    \label{eq:LB}
\end{equation}
where  $\gamma_{\tc,k}$ and $\gamma_{\tp,,k}$ denote, respectively, the effective common and private signal-to-interference-and-noise-ratios (SINRs) of user $k$ and are given by (see \cite{Medard})
\begin{equation}
    \begin{aligned}
    \gamma_{\tc,k}&=\frac{|\hat{b}_{\tc,k}|^2}{\text{var}(\eta_{\tc,k})} &&=\frac{|\mathbb{E}[\bs{h}_k^\HH \bs{p}_\tc]|^2}{\text{var}(\bs{h}_k^\HH\bs{p}_\tc)+\sum_{i=1}^K \mathbb{E}[|\bs{h}_k^\HH \bs{p}_{\tp,i}|^2] + 1}\\
    \gamma_{\tp,k}&=\frac{|\hat{b}_{\tp,k,k}|^2}{\text{var}(\eta_{\tp,k})} &&=\frac{|\mathbb{E}[\bs{h}_k^\HH \bs{p}_{\tp,k}]|^2}{\text{var}(\bs{h}_k^\HH\bs{p}_{\tp,k})+\sum_{\substack{i=1\\i\neq k}}^K \mathbb{E}[|\bs{h}_k^\HH \bs{p}_{\tp,i}|^2] + 1}.
    \end{aligned}
    \label{eq:SINRs}
\end{equation}
After inserting the expressions for the bilinear precoders [see \eqref{eq:pc} and \eqref{eq:pk}] into \eqref{eq:SINRs}, we obtain the following common and private SINRs of user $k$
\begin{align}
    \gamma_{\tc,k}&=\frac{|\bs{z}_k^\HH \bs{a}_\tc|^2}{\bs{a}_\tc^\HH \bs{Z}_k \bs{a}_\tc+|\boldsymbol{q}_k^\HH  \boldsymbol{a}_{\text{p},k}|^2+\sum_{i=1}^K \bs{a}_{\tp,i}^\HH \bs{Q}_{i,k} \bs{a}_{\tp,i} +1}\label{eq:SINRc}\\
    \gamma_{\tp,k}&=\frac{|\bs{q}_k^\HH  \bs{a}_{\tp,k}|^2}{\sum_{i=1}^K \bs{a}_{\tp,i}^\HH \bs{Q}_{i,k} \bs{a}_{\tp,i} +1 }. \label{eq:SINRp}
\end{align}
A detailed derivation of both closed-form expressions is presented in Appendix~\ref{app1}, where the definitions of the $\bs{z}_k$, $\bs{Z}_k$ $\bs{Q}_{i,k}$ and $\bs{q}_k$ are given. Note that $\bs{a}_{\tc}=\text{vec}(\bs{A}_{\tc})$ and $\bs{a}_{\tp,i}=\text{vec}(\bs{A}_{\tp,i})$ [see \eqref{eq:apk}].
\par Now that we derived the SINR expressions, we can formulate sum rate optimization problem under the transmit power constraints as follows
\begin{equation}
    \begin{aligned}
    \underset{\substack{\alpha_\tc \in [0,1]\\ R_\tc,\bs{p}_\tc,\bs{p}_{\tp,1}, \dots, \bs{p}_{\tp,K}}}{\max} \: R_\tc+\sum_{k=1}^K R_{\tp,k}^\text{LB} \quad  \text{s.t.} \quad R_\tc &\leq R_{\tc,k}^\text{LB}, \forall k\\
    \sum_{k=1}^K \mathbb{E}[\|\bs{p}_{\tp,k}\|^2]&\leq (1-\alpha_\tc) P_\text{dl}\\
     \mathbb{E}[\|\bs{p}_\tc\|^2]&\leq \alpha_\tc P_\text{dl}
    \end{aligned}
    \label{eq:OptPrimal}
\end{equation}
where the first constraint ensures that the common rate is not larger than the minimum common rate of all users, which allows the common message to be be decoded by all users. \\
The optimization problem in \eqref{eq:OptPrimal} is non-convex and hence intractable. To reduce the computational cost, we propose a solution approach similar to \cite{Kaulich}, which mainly consists of splitting \eqref{eq:OptPrimal} into three parts.
\begin{itemize}
    \item[(I)] First, assuming a fixed power fraction $\alpha_\tc$, we formulate the private rate optimization problem 
\begin{equation}
    \underset{\bs{p}_{\tp,1}, \dots, \bs{p}_{\tp,K}}{\max} \: \sum_{k=1}^K R_{\tp,k}^\text{LB} \quad \text{s.t.} \quad
    \sum_{k=1}^K \mathbb{E}[\|\bs{p}_{\tp,k}\|^2]\leq (1-\alpha_\tc) P_\text{dl}.
    \label{eq:PriRateOpt}
\end{equation}
Here, we implicitly assume that the private precoder optimization does not affect the common rate, which is of course not the case in general, since the common SINR depends on the private precoders as well. But this assumption eases the solving procedure by decoupling the private and common rate optimization. Note that this assumption is only considered to make the optimization problem \eqref{eq:OptPrimal} more tractable. The system model and the SINR expressions for the performance evaluation do take into account the effect of the private precoders on the common rate.
\item[(II)] Now, for fixed private precoders and some given power fraction $\alpha_\tc$, we recast the common precoder optimization problem as a max-min beamforming problem
\begin{equation}
    \underset{\bs{p}_\tc}{\max} \: \underset{k}{\min} \: R_{\tc,k}^\text{LB} \quad \text{s.t.} \quad \mathbb{E}[\|\bs{p}_\tc\|^2]\leq \alpha_\tc P_\text{dl}.
    \label{eq:ComRateOpt1}
\end{equation}
Since the rate expression $R_{\tc,k}^\text{LB}$ is monotonically increasing in $\gamma_{\tc,k}$, we can reformulate \eqref{eq:ComRateOpt1} as a max-min problem of the common SINRs, that is,
\begin{equation}
    \underset{\bs{p}_\tc}{\max} \: \underset{k}{\min} \: \gamma_{\tc,k} \quad \text{s.t.} \quad \mathbb{E}[\|\bs{p}_\tc\|^2]\leq \alpha_\tc P_\text{dl}.
    \label{eq:ComRateOpt2}
\end{equation}
\item[(III)] We are now left with the optimization over the power fraction $\alpha_\tc$ that is allocated to the common stream transmission. Finding an optimal power fraction would require a joint optimization over the precoders such that the overall sum rate is maximized. In order to circumvent the complexity of such an approach, we opt for a line search method that returns the scalar $\alpha_\tc$ that maximizes the overall rate. 
\end{itemize}

\section{Solution Approach}
\label{sec3}
In this section, we present a detailed description of our solution approach. Our contribution here consists of adapting and combining several methods, which were shown to have a satisfactory performance for a different setup than ours. As mentioned in the previous section, we divide the sum rate optimization problem into 3 subproblems consisting of an outer loop for the power fraction optimization and an inner loop with 2 optimization steps: the private precoder optimization and the common precoder optimization.
\par For the private precoder optimization, we use an efficient method which was proposed in \cite{FracProg} and used in \cite{WSRM} to solve the weighted sum rate maximization problem. The method makes use of the so-called \textit{Lagrangian dual transform} and \textit{quadratic transform} \cite{FracProg}. Both transforms consist of introducing auxiliary variables to reformulate the underlying optimization problem. On the one hand, it is ensured such that the newly formulated problem is equivalent to the original one in the optimum. On the second hand, the solving procedure of the new optimization problem becomes much easier. This is because each variable can be computed in closed-form when the remaining ones are fixed and one needs to alternate between the different optimization variables until convergence is reached.
\par As for the common precoder optimization, we again exploit the quadratic transform \cite{FracProg} to reformulate the underlying maximization problem by introducing some auxiliary variables. To solve the so-obtained optimization problem, we propose two different approaches. One is to use an interior-point solver, e.g., CVX \cite{cvx}. The other approach is an iterative algorithm inspired by \cite{Hunger}. It consists of iterativetly updating the common transformation vector, while ensuring that the minimum common SINR does not decrease after each update, which gives rise to the name \textit{"Common SINR increasing algorithm"}.
\par For both, the private and the common precoder optimization problems, we assumed that the power fraction allocated to the common stream is fixed. Now, in order to find the power fraction that maximizes the overall sum rate, we consider an outer loop for the power fraction optimization and perform a line search in the interval $[0,1]$. To this end, we use a simple and efficient approach for finding the maximum of a function, namely the golden section method \cite{GoldenSection}. The inner loop consists thus of the first two optimization problems stated before: the private precoder optimization and the common precoder optimization.
\subsection{(I) Private Precoder Optimization} \label{sec:PriPreOpt}
Before we proceed to the solution approach for the private precoder optimization problem, we rewrite \eqref{eq:PriRateOpt} as a function of $\bs{a}_\tp=[\bs{a}_{\tp,1}^\mathrm{T}, \dots, \bs{a}_{\tp,K}^\mathrm{T}]^\mathrm{T}$ containing the vectorized forms of the transformation matrices $\bs{A}_{\tp,1},\dots,\bs{A}_{\tp,K}$ of the private precoders 
\begin{equation}
    \underset{\bs{a}_{\tp}}{\max} \: \sum_{k=1}^K \log_2(1+\gamma_{\tp,k}) \quad \text{s.t.} \quad\,
    \bs{a}_{\tp}^\HH \bs{F} \bs{a}_{\tp} \leq (1-\alpha_\tc) P_\text{dl}
    \label{eq:PriRateOpt2}
\end{equation}
where we used the reformulation of the power constraint in \eqref{eq:PriPowCons} and applied the identities in \eqref{eq:TrVec} resulting in the following expression
\begin{equation}
\sum_{i=1}^K \bs{a}_{\tp,i}^\HH \bs{F}_{i} \bs{a}_{\tp,i}\leq (1-\alpha_\tc) P_\text{dl}.
\end{equation}
The matrices $\bs{F}_{i}$  and the matrix $\bs{F}$ are given by 
\begin{align}
    &\bs{F}_{i}=\bs{C}_{\bs{y}_i}^\mathrm{T} \otimes \bs{I}_M\\
    &\bs{F}=\text{blkdiag}\left(\bs{F}_{1}, \bs{F}_{2},\dots, \bs{F}_{K}\right) .
\end{align}
\par Due to the sum of the logarithms and the dependence of the private SINR of each user on the transformation matrices of all users, the optimization problem in \eqref{eq:PriRateOpt2} is non-convex and therefore intractable. To address this challenge, we proceed by decoupling the optimization variables in the objective function and splitting \eqref{eq:PriRateOpt2} into several subproblems, that are much easier to solve. To this end, we employ the Lagrangian dual transform \cite{FracProg}.\\
The first step of this transform is to introduce a vector of slack variables $\bs{\alpha}=[\alpha_1,\dots,\alpha_K]$ and formulate the following optimization problem 
\begin{equation}
    \underset{\bs{a}_{\tp}, \bs{\alpha}}{\max} \:  f_{1a}(\bs{a}_{\tp}, \bs{\alpha}) \quad \text{s.t.} \quad\,
    \bs{a}_{\tp}^\HH \bs{F} \bs{a}_{\tp} \leq (1-\alpha_\tc) P_\text{dl}
    \label{eq:PriRateOpt1a}
\end{equation}
where the new objective function for the private precoder optimization is given by
\begin{equation}
    f_{1a}(\bs{a}_{\tp}, \bs{\alpha})=\sum_{k=1}^{K} \log_2(1+\alpha_k)- \sum_{k=1}^{K} \alpha_k + \sum_{k=1}^{K} \frac{(1+\alpha_k)\gamma_{\tp,k}}{1+\gamma_{\tp,k}}.
    \label{eq:1a}
\end{equation}
In this formulation of the objective function, the private SINR $\gamma_{\tp,k}$ lies outside of the logarithm function, which eases the optimization with respect to the transformations $\bs{a}_{\tp,k}$ as we shall see in the upcoming steps. \\
One further observation is that for fixed $\bs{a}_{\tp}$, the optimal $\alpha_k$ is given by
\begin{equation}
    \alpha_k^\text{opt}=\gamma_{\tp,k}.
    \label{eq:alphaK}
\end{equation}
The equivalence of \eqref{eq:PriRateOpt2} and \eqref{eq:PriRateOpt1a} can be seen by inserting \eqref{eq:alphaK} into \eqref{eq:1a}, which leads to the same objective of \eqref{eq:PriRateOpt2}.
\par If we now fix $\alpha_k$, optimizing over $\bs{a}_{\tp}$ reduces to maximizing the third term of \eqref{eq:1a} while taking into account the transmit power constraint. The optimization problem therefore reads as
\begin{equation}
     \underset{\bs{a}_{\tp}}{\max} \: f_2(\bs{a}_{\tp}) \quad \text{s.t.} \quad \bs{a}_{\tp}^\HH \bs{F} \bs{a}_{\tp} \leq (1-\alpha_\tc) P_\text{dl}
     \label{eq:a}
\end{equation}
where the objective function is given by
\begin{equation}
         f_2(\bs{a}_{\tp})=\sum_{k=1}^{K}  \frac{(1+\alpha_k)|\bs{q}_k^\HH  \bs{a}_{\tp,k}|^2}{1+ \sum_{i=1}^K \bs{a}_{\tp,i}^\HH \bs{Q}_{i,k} \bs{a}_{\tp,i} +|\bs{q}_k^\HH  \bs{a}_{\tp,k}|^2}
    \label{eq:f2}
\end{equation}
after incorporating \eqref{eq:SINRp}.
The optimization problem in \eqref{eq:a} is a multi-ratio fractional program, to which we apply the quadratic transform as proposed in \cite{FracProg}. This transform allows to decouple the numerators and denominators of the individual fractions while guaranteeing the equivalence of the transformed problem to the original one.
\par In order to apply the quadratic transform, we need to introduce an auxiliary variable $\beta_k$ for each of the $K$ ratios in \eqref{eq:f2}. To this end, we formulate the following optimization problem
\begin{equation}
    \underset{\bs{a}_{\tp},\bs{\beta}}{\max} \: f_{2a}(\bs{a}_{\tp},\bs{\beta}) \quad \text{s.t.} \quad \bs{a}_{\tp}^\HH \bs{F} \bs{a}_{\tp} \leq (1-\alpha_\tc) P_\text{dl}
     \label{eq:2a}
\end{equation}
where $\bs{\beta}=[\beta_1,\beta_2,\dots,\beta_K]$ and the objective function $f_{2a}$ is given by
\begin{align}
    f_{2a}(\bs{a}_{\tp},\bs{\beta})&=\sum_{k=1}^{K} 2\sqrt{1+\alpha_k}\Re\{\beta_k^*\bs{q}_k^\HH  \bs{a}_{\tp,k} \} \nonumber\\
    &- \sum_{k=1}^{K} |\beta_k|^2 \left(|\bs{q}_k^\HH  \bs{a}_{\tp,k}|^2 + \sum_{i=1}^K \bs{a}_{\tp,i}^\HH \bs{Q}_{i,k} \bs{a}_{\tp,i} +1  \right).
\end{align}
The quadratic transform guarantees that solving \eqref{eq:2a} for $\bs{\beta}$ and $\bs{a}_{\tp}$ is equivalent to solving \eqref{eq:a}.
\par The optimization problem in \eqref{eq:2a} is convex in each of the variables $\bs{\beta}$ and $\bs{a}_{\tp}$, when the other one is fixed. A straightforward solution approach is to alternately solve for one of the variables while maintaining the other one constant until convergence of the objective is reached. Here, convergence is attained, when the absolute change of the objective function drops below some threshold $\varepsilon$.
\par Starting by fixing $\bs{a}_{\tp}$, the optimal $k$th element in $\bs{\beta}$ that maximizes the objective in \eqref{eq:2a} results from setting the derivative of $f_{2a}$ with respect to $\beta_k$ to zero (the power constraint is independent of the auxiliary variable $\beta_k$) and is given by 
\begin{equation}
    \beta_k^\text{opt}=\frac{\sqrt{1+\alpha_k}\bs{q}_k^\HH  \bs{a}_{\tp,k}}{1+ \sum_{i=1}^K \bs{a}_{\tp,i}^\HH \bs{Q}_{i,k} \bs{a}_{\tp,i} +|\bs{q}_k^\HH  \bs{a}_{\tp,k}|^2}.
    \label{eq:betaK}
\end{equation}
\par The second step is to fix $\bs{\beta}$ and solve for the transformations $\bs{a}_{\tp,k}$. In this case, \eqref{eq:2a} is convex in $\bs{a}_{\tp,k}$ and the Karush-Kuhn-Tucker (KKT) conditions are not only necessary but also sufficient for global optimality. Hence, based on setting the derivative of the Lagrangian function corresponding to \eqref{eq:2a} w.r.t. $\bs{a}_{\tp,k}$ to zero, we obtain
\begin{align}
    \bs{a}_{\tp,k}(\lambda)=&\sqrt{1+\alpha_k} \beta_k \nonumber\\
    &\left( |\beta_k|^2 \bs{q}_k \bs{q}_k^\HH + \sum_{i=1}^K |\beta_i|^2 \bs{Q}_{k,i} + \lambda \bs{F}_{k}\right) ^{-1} \bs{q}_k.
    \label{eq:akLambda}
 \end{align}
 with the Lagrangian multiplier $\lambda\geq 0$ associated to the transmit power constraint.
 The optimal Lagrangian dual variable $\lambda^\text{opt}$ can be obtained via bisection such that the transmit power constraint is fulfilled with a certain accuracy $\epsilon>0$. Note that $ \bs{a}_{\tp,k}(\lambda)$ is monotonically decreasing in $\lambda$, therefore the function defining the power constraint $\sum_{i=1}^K \bs{a}_{\tp,i}^\HH(\lambda) \bs{F}_{i} \bs{a}_{\tp,i}(\lambda)$ is also monotonically decreasing in $\lambda$. Hence, searching for the optimal $\lambda$ that satisfies the power constraint corresponds to finding the zero of the monotonic function $\sum_{i=1}^K \bs{a}_{\tp,i}^\HH(\lambda) \bs{F}_{i} \bs{a}_{\tp,i}(\lambda)-(1-\alpha_\tc)P_\text{dl}$ \cite{bisec}.
\par The solution approach to the private precoder optimization problem \eqref{eq:PriRateOpt2} is given in Algorithm~\ref{alg:PrivatePrec}.\\
 \begin{algorithm}
\caption{Private Precoder Optimization Algorithm}\label{alg:PrivatePrec}
\begin{algorithmic}[1]
\State Initialize $\bs{a}_{\tp}$% 
\State Compute $\alpha_k$ and $\beta_k$ according to \eqref{eq:alphaK} and \eqref{eq:betaK}
\State Search for the optimal Lagrangian dual variable $\lambda^\text{opt}$ using bisection 
\State Update the transformation vectors $\bs{a}_{\tp,k}$ according to \eqref{eq:akLambda}
\State Iterate between Steps 2--4 until the objective function $f_{1a}$ convergences to a local optimum
\end{algorithmic}
\end{algorithm}
 \subsection{(II) Common Precoder Optimization}\label{sec:ComPreOpt}
 To tackle the max-min beamforming problem stated in \eqref{eq:ComRateOpt2}, we apply the quadratic transform  employed in the previous section and propose two approaches to solve the reformulated optimization problem.
\par First, we express \eqref{eq:ComRateOpt2} depending on the common transformation vector $\bs{a}_\tc$ 
 \begin{equation}
     \underset{\bs{a}_\tc}{\max}\: \underset{k}{\min} \: \gamma_{\tc,k} \quad \text{s.t.} \quad \bs{a}_\tc^\HH \bs{F} \bs{a}_\tc \leq \alpha_\tc P_\text{dl}
 \end{equation}
 where we performed similar manipulations of the power constraint in \eqref{eq:ComPowCons} as in the previous section, so that the matrix $\bs{F}$ is again given by
 \begin{equation}
     \bs{F}=\text{blkdiag}\left(\bs{F}_{1}, \bs{F}_{2},\dots, \bs{F}_{K}\right)=\bs{C}_{\bs{y}}^\mathrm{T} \otimes \bs{I}_{MT_\text{dl}}.
 \end{equation}
 \par A common strategy to circumvent the max-min optimization is to introduce a slack variable $z$ such that the minimization with respect to the user index is recast as a constraint and where the resulting optimization problem is equivalent to the original max-min problem. Introducing the slack variable $z$ yields the following optimization problem
 \begin{equation}
    \begin{aligned}
     \underset{\bs{a}_\tc,z}{\max} \: z \quad 
     \text{s.t.} \quad & \bs{a}_\tc^\HH \bs{F} \bs{a}_\tc \leq \alpha_\tc P_\text{dl}\\
     & \gamma_{\tc,k}\geq z, \forall k.
     \end{aligned}
     \label{eq:ComOpt2}
 \end{equation}
 \par Since $\gamma_{\tc,k}$, the common SINR of the $k$th user, is a fraction where the numerator and denominator depend on the optimization variable $\bs{a}_\tc$, we use the quadratic transform in order to decouple the numerator and the denominator.
 \par By defining an auxiliary variable $\eta_k$ for each of the $K$ common SINRs, where $\bs{\eta}=[\eta_1,\dots,\eta_K]$, we can reformulate an optimization equivalent to problem \eqref{eq:ComOpt2}
\begin{equation}
    \begin{aligned}
    \underset{\bs{a}_\tc,z,\bs{\eta}}{\max} \: z \quad 
     \text{s.t.} \quad & \bs{a}_\tc^\HH \bs{F} \bs{a}_\tc \leq \alpha_\tc P_\text{dl}\\
     &2\Re\{\eta_k^* \bs{z}_k^\HH \bs{a}_\tc \} - |\eta_k|^2 (\bs{a}_\tc^\HH \bs{Z}_{k} \bs{a}_\tc + \sigma_k^2 ) \geq z, \forall k
    \end{aligned}
    \label{eq:ComOpt3}
\end{equation} 
where we defined $\sigma_k^2=|\boldsymbol{q}_k^\HH  \boldsymbol{a}_{\text{p},k}|^2+\sum_{i=1}^{K} \bs{a}_{\tp,i}^\HH \bs{Q}_{i,k} \bs{a}_{\tp,i} +1$, which obviously does not depend on the common transformation vector $\bs{a}_\tc$.
\par For fixed $\bs{a}_\tc$, the optimal auxiliary variables $\eta_k^\text{opt}$ can be determined by setting the derivative of the left-hand side of the SINR constraint in \eqref{eq:ComOpt3} to zero, which yields the following result
\begin{equation}
    \eta_k^\text{opt}=\frac{ \bs{z}_k^\HH \bs{a}_\tc}{\bs{a}_\tc^\HH \bs{Z}_k \bs{a}_\tc + \sigma_k^2}.
    \label{eq:etaOpt}
\end{equation}
We note that plugging this expression into \eqref{eq:ComOpt3} leads to the same optimization problem in \eqref{eq:ComOpt2}.\\
If we now fix the auxiliary variables  $\eta_k$, we end up with a convex optimization problem, which can be solved via CVX \cite{cvx}.
\par Alternatively, we propose an iterative approach inspired by \cite{Hunger} that aims to increase the common SINR in each of its iterations.\\
Consider again the optimization problem in \eqref{eq:ComOpt3} where the auxiliary variables $\eta_k$ are kept constant. If we denote by $\ell(n)$ the user with the least common SINR after iteration $n$, the update rule for the common transformation vector $\bs{a}_\tc$ at iteration $n+1$ reads as
\begin{equation}
    \bs{a}_\tc^{(n+1)}=(1-u)\bs{a}_\tc^{(n)} + v \bs{Z}_{\ell(n)}^{-\frac{1}{2}} \bs{a}_\tc^{(n),\perp}
    \label{eq:AcUpdate}
\end{equation}
where $u\in[0,1]$ denotes the step-size and hence controls how much from the previous update is included in the current update and $\bs{Z}_k$ is defined in \eqref{eq:Zk}. The length of the step size is adapted during the algorithm to speed up the convergence.\\
The second term in \eqref{eq:AcUpdate} involves different variables. $v$ is a complex scalar that we have to tune together with the vector $\bs{a}_\tc^{(n),\perp}$ for the update in \eqref{eq:AcUpdate} to ensure that the minimum common SINR at least does not decrease in step $(n+1)$ compared to step $(n)$ and that the power constraint is satisfied. Applying $\bs{Z}_{\ell(n)}^{-\frac{1}{2}}$ (the inverse of the square root matrix of $\bs{Z}_k=\bs{Z}_{\ell(n)}^{\frac{1}{2},\HH}\bs{Z}_{\ell(n)}^{\frac{1}{2}}$) to $\bs{a}_\tc^{(n),\perp}$ simplifies the tuning of the parameters $v$ and $\bs{a}_\tc^{(n),\perp}$ as it is presented in Appendix~\ref{app2}.\\
Together with the conditions implied by the power constraint, the parameters $v$ and $\bs{a}_\tc^{(n),\perp}$ are given by
\begin{align}
    & v=\sqrt{\alpha_\tc P_\text{dl}(2u-u^2)} \mathrm{e}^{-\mathrm{j}\angle\left(\bs{t}_\tc^{(n),\HH}\bs{a}_\tc^{(n),\perp}\right)}\label{eq:v}\\
    & \bs{a}_\tc^{(n),\perp,\prime}= \bs{T}^{(n)} \bs{t}_\tc^{(n)} \label{eq:acnpp}\\
    & \bs{a}_\tc^{(n),\perp} = \frac{\bs{a}_\tc^{(n),\perp,\prime}}{\|\bs{F}^{\frac{1}{2}}\bs{Z}_{\ell(n)}^{-\frac{1}{2}}\bs{a}_\tc^{(n),\perp, \prime}\|_2} \label{eq:acnp}
\end{align}
where $\bs{t}_\tc^{(n)}=\eta_{\ell(n)}\bs{Z}_{\ell(n)}^{-\frac{1}{2},\HH}\bs{z}_{\ell(n)}-|\eta_{\ell(n)}|^2(1-u) \bs{Z}_{\ell(n)}^{\frac{1}{2}}\bs{a}_\tc^{(n)}$ as defined in Appendix~\ref{app2}.\\
The matrix $\bs{T}^{(n)}$ in \eqref{eq:acnpp} is the orthogonal projector onto the nullspace of the vector $\bs{t}^{(n)}=\bs{Z}_{\ell(n)}^{-\frac{1}{2},\HH }  \bs{F}^\HH \bs{a}_\tc^{(n)}$ and is given by
\begin{equation}
    \bs{T}^{(n)}= \bs{I} - \frac{\bs{t}^{(n)}\bs{t}^{(n),\HH}}{\|\bs{t}^{(n)}\|_2^2}.
\end{equation}
\par Suppose that $\ell(n)$ is the user with the least SINR at the beginning of iteration $n$. Then, we compute the parameters $v$ and $\bs{a}_\tc^{(n),\perp}$ according to \eqref{eq:v} and \eqref{eq:acnp}, respectively. Using these parameters, we can find the temporary update $\bs{a}_\tc^\text{temp}$ using the current step size. This update might lead to an SINR of some other user $k$ smaller than $\gamma_\tc^\text{min}=\gamma_{\tc,\ell(n)}$. If this is the case, then the update is not successful and the step-size is halved. Otherwise, the latter is increased by a factor of 2 and the update is performed as long as it does not drop the resulting minimum common SINR below $\gamma_\tc^\text{min}$.\\
The common precoder optimization via the iterative approach presented above can be summarized in Algorithm~\ref{alg:IncSINR1}. Note that Algorithm~\ref{alg:IncSINR1} calls Algorithm~\ref{alg:IncSINR2} in step 3.
\begin{algorithm}
\caption{Common Precoder Optimization Algorithm}\label{alg:IncSINR1}
\begin{algorithmic}[1]
\State Initialize: $\bs{a}_\tc$ 
\State Compute $\eta_k$ according to \eqref{eq:etaOpt}
\State Run the iterative common SINR increasing algorithm presented in Algorithm~\ref{alg:IncSINR2}
\State Alternate between Step 2 and 3 until convergence is reached
\end{algorithmic}
\end{algorithm}

\begin{algorithm}
\caption{Common SINR Increasing Algorithm}\label{alg:IncSINR2}
\begin{algorithmic}[1]
\State Initialize: $u=u_\text{max}$, $n=1$ 
\While{$n\leq N_\text{max}$}
\State $n\leftarrow n+1$
\State $\ell(n)=\underset{k}{\text{arg}\min}\:\gamma_{\tc,k}$ [cf. \eqref{eq:SINRc}]
\State $\gamma_\tc^\text{min}=\gamma_{\tc,\ell(n)}$
\State Compute $\bs{a}_\tc^{(n),\perp}$ according to \eqref{eq:acnp}
\State Compute $v$ according to \eqref{eq:v}
\State Find $\bs{a}_\tc^\text{temp}=(1-u)\bs{a}_\tc^{(n)}+v \bs{a}_\tc^{(n),\perp} $
\State Calculate the temporary minimum common SINR $\gamma_\tc^\text{temp}$
\If{$\gamma_\tc^\text{temp}>\gamma_\tc^\text{min}$}
\State $\bs{a}_\tc^{(n+1)}=\bs{a}_\tc^\text{temp}$
\If{$u\leq u_\text{max}$}
\State $u\leftarrow 2u$
\EndIf
\Else
\State $u\leftarrow u/2$
\EndIf
\EndWhile
\end{algorithmic}
\end{algorithm}
 \subsection{(III) Power Fraction Optimization}
 In order to find $\alpha_\tc \in [0,1]$, the power fraction allocated to the common stream, we use a one-dimensional search approach, known as the golden section method. It consists of finding the maximum/minimum of a certain function over some interval $[a,b]$ by efficiently shrinking the search interval after each step \cite{GoldenSection}. 
 \par For our setup, the function of interest is the sum rate $R_\text{sum}^\text{LB,RS}=R_\tc+\sum_{k=1}^K R_{p,k}^\text{LB}$, i.e., the objective of \eqref{eq:OptPrimal}. The search interval is given by $[a,b]=[0,1]$ and the method operates as follows. First, two intermediate values $\alpha_{\tc1}$ and $\alpha_{\tc2}$ are found according to 
 \begin{align}
     \alpha_{\tc1}= a + (1-g) (b-a) \label{eq:x1}\\
     \alpha_{\tc2}= a + g (b-a)\label{eq:x2}
 \end{align}
 where $g$ is the inverse of the golden ratio $\frac{\sqrt{5}+1}{2}$ which gives rise to the name of this method.
 \par Now, for the given $\alpha_{\tc1}$ and $\alpha_{\tc2}$, the two potential values of $\alpha_\tc$, we run the private precoder optimization approach described in Sec.~\ref{sec:PriPreOpt} for a fixed common precoder. We then proceed with solving the common precoder optimization problem following the approach presented in Sec.~\ref{sec:ComPreOpt}.\\
 We hence obtain two values for the sum rate $R_\text{sum,1}^\text{LB,RS}=R^\text{sum}(\alpha_{\tc1})$ and $R_\text{sum,2}^\text{LB,RS}=R^\text{sum}(\alpha_{\tc2})$ corresponding to $\alpha_{\tc1}$ and $\alpha_{\tc2}$. \\
 If $R_\text{sum,1}^\text{LB,RS}>R_\text{sum,2}^\text{LB,RS}$, then the search interval is reduced and the new boundaries are given by $a$ and $b=\alpha_{\tc2}$, otherwise the new interval boundaries are $a=\alpha_{\tc1}$ and $b$. \\
 Given the new search interval, we can recompute $\alpha_{\tc1}$ and $\alpha_{\tc2}$ according to \eqref{eq:x1} and \eqref{eq:x2} and then go through the previously described steps until the size of the interval given by $|b-a|$ drops below a certain threshold or the maximum number of iterations is reached.

\section{Iterative Weighted MSE Approach}
\label{sec4}
To compare the performance of the proposed approach, we briefly present the iterative weighted minimum mean squared error (IWMMSE) precoding approach proposed in \cite{IWMMSE}, \cite{Clerckx_RS2}. This method is based on establishing a relationship between the sum rate maximization problem and the augmented weighted MSE minimization problem by introducing weights and equalizing filters. The latter optimization problem turns out to be convex in each of its variables when the other ones are kept constant and hence an alternating solution approach is pursued. 
\par Let $g_{\tc,k}$ and $g_{\tp,k}$ be the equalizing filters for the $k$th user's common and private messages, respectively. 
Given the expressions in \eqref{eq:rck} and in \eqref{eq:rpk}, the estimates for the $k$th user common message and private message read as
\begin{align}
    \hat{s}_{\tc,k}&=g_{\tc,k} r_{\tc,k}\nonumber\\
    &= g_{\tc,k} \bs{h}_k^\HH \bs{p}_\tc s_\tc + g_{\tc,k} \sum_{j=1}^{K} \bs{h}_k^\HH \bs{p}_{\tp,j} s_{\tp,j} + g_{\tc,k} v_k\\
    \hat{s}_{\tp,k}&=g_{\tp,k}  r_{\tp,k} \nonumber\\
    &= g_{\tp,k} \sum_{j=1}^{K} \bs{h}_k^\HH \bs{p}_{\tp,j} s_{\tp,j} + g_{\tp,k} v_k
\end{align}
respectively.
\par The MSEs of the common message and the private message for user $k$ can respectively be written as
\begin{align}
    &\varepsilon_{\tc,k}=\mathbb{E}[|\hat{s}_{\tc,k}-s_\tc|^2]=\nonumber\\
    &|g_{\tc,k}|^2(|\bs{h}_k^\HH \bs{p}_\tc|^2 + \sum_{j=1}^{K}  |\bs{h}_k^\HH \bs{p}_j|^2 + 1 ) + 1 -2 \Re\{g_{\tc,k}^* \bs{p}_\tc^\HH \bs{h}_k\}\\
     &\varepsilon_{\tp,k}=\mathbb{E}[|\hat{s}_{\tp,k}-s_{\tp,k}|^2]=\nonumber\\
     &|g_{\tp,k}|^2( \sum_{j=1}^{K}  |\bs{h}_k^\HH \bs{p}_{\tp,j}|^2 + 1 ) + 1 -2 \Re\{g_{\tp,k}^* \bs{p}_{\tp,k}^\HH \bs{h}_k\}.
\end{align}
Minimizing the MSEs with respect to the equalizing filters $g_{\tc,k}$ and $g_{\tp,k}$, yields the MMSE optimal filters 
\begin{equation}
    g_{\tc,k}^\text{MMSE}=\frac{\bs{p}_\tc^\HH \bs{h}_k}{T_{\tc,k}}, \qquad
     g_{\tp,k}^\text{MMSE}=\frac{\bs{p}_{\tp,k}^\HH \bs{h}_k}{T_{\tp,k}}
    \label{eq:gckgk}
\end{equation}
respectively, where $T_{\tc,k}= |\bs{h}_k^\HH \bs{p}_\tc|^2 + \sum_{j=1}^{K}  |\bs{h}_k^\HH \bs{p}_{\tp,j}|^2 + 1 $ and $T_{\tp,k}=\sum_{j=1}^{K}  |\bs{h}_k^\HH \bs{p}_{\tp,j}|^2 + 1 $.\\
Applying the MMSE equalizing filters from \eqref{eq:gckgk} yields the following MMSE expressions, which are closely related to the instantaneous common and private SINRs
\begin{align}
   &\varepsilon_{\tc,k}^\text{MMSE}=1-\frac{|\bs{h}_k^\HH \bs{p}_\tc|^2}{T_{\tc,k}}=\frac{1}{1+\gamma_{\tc,k}^\text{inst}} \\
   &\varepsilon_{\tp,k}^\text{MMSE}=1-\frac{|\bs{h}_k^\HH \bs{p}_{\tp,k}|^2}{T_{\tp,k}}=\frac{1}{1+\gamma_{\tp,k}^\text{inst}}
    \label{eq:MMSEs}
\end{align}
where 
\begin{align*}
    \gamma_{\tc,k}^\text{inst}&=\frac{|\bs{h}_k^\HH \bs{p}_\tc|^2}{\sum_{j=1}^{K}  |\bs{h}_k^\HH \bs{p}_{\tp,j}|^2 + 1}\\
    \gamma_{\tp,k}^\text{inst}&=\frac{|\bs{h}_k^\HH \bs{p}_{\tp,k}|^2}{\sum_{j\neq k}  |\bs{h}_k^\HH \bs{p}_{\tp,j}|^2 + 1}
\end{align*}
\par Now, introducing the common and private weights $u_{\tc,k}$ and $u_{\tp,k}$, we consider the augmented weighted MSEs
\begin{equation}
 \begin{aligned}
    &\xi_{\tc,k}=u_{\tc,k}\varepsilon_{\tc,k} - \log_2(u_{\tc,k})\\
    &\xi_{\tp,k}=u_{\tp,k}\varepsilon_{\tp,k} - \log_2(u_{\tp,k}).
\end{aligned}   
\label{eq:xick}
\end{equation}
When the MSEs are fixed, the optimal weights that minimize $\xi_{\tc,k}$ and $\xi_{\tp,k}$ are given by $u_{\tc,k}^*=1/\varepsilon_{\tc,k}$ and $u_{\tp,k}^*=1/\varepsilon_{\tp,k}$ (a scaling factor of $1/\log(2)$ was ignored, since it does not affect the solution). \\
Note that inserting these expressions into \eqref{eq:xick} and applying the MMSE equalizing filters yields the following augmented weighted MMSEs
\begin{equation}
    \xi_{\tc,k}^\text{MMSE}=1-R_{\tc,k}^\text{inst}, \quad  \xi_{\tp,k}^\text{MMSE}=1-R_{\tp,k}^\text{inst} \label{eq:XiRate}
\end{equation}
where $R_{\tc,k}^\text{inst}$ and $R_{\tp,k}^\text{inst}$ are the common and private rates of user $k$, respectively, given by $R_{\tc,k}^\text{inst}=\log_2(1+\gamma_{\tc,k}^\text{inst})$ and $R_{\tp,k}^\text{inst}=\log_2(1+\gamma_{\tp,k}^\text{inst})$.
\par Since the BS has only access to an observation of the channel from which the estimate $\hat{\bs{H}}$ can be computed, the instantaneous achievable rates are not known to the BS. In \cite{Clerckx_RS2}, the authors thus define the conditional average common and private rates of user $k$, namely $\bar{R}_{\tc,k}=\mathbb{E}[R_{\tc,k}^\text{inst}|\hat{\bs{H}}]$ and $\bar{R}_{\tp,k}=\mathbb{E}[R_{\tp,k}^\text{inst}|\hat{\bs{H}}]$, to which the BS has access. The average common rate is then given by $\bar{R}_\tc=\underset{k}{\min}\: \bar{R}_{\tc,k}$. \\
The average sum rate maximization problem hence reads as
\begin{equation}
    \begin{aligned}
    \underset{\substack{\bar{R}_\tc,\bs{p}_\tc\\ \bs{p}_{\tp,1},\dots, \bs{p}_{\tp,K}}}{\max} \: &\bar{R}_\tc + \sum_{k=1}^K  \bar{R}_{\tp,k} \\
    &\text{s.t.} \quad \bar{R}_\tc\leq \bar{R}_{\tc,k}, \forall k\\
    &\|\bs{p}_\tc\|^2+\sum_{k=1}^K \|\bs{p}_{\tp,k}\|^2 \leq P_\text{dl}.
    \end{aligned}
    \label{eq:ASR}
\end{equation}
Note that contrarily to \eqref{eq:OptPrimal}, the optimization problem \eqref{eq:ASR} exhibits an instantaneous power constraint and has to be solved in every channel coherence interval. 
\par Due to the intractability of \eqref{eq:ASR}, the so-called sample average approximation (SAA) is used in order to evaluate the average rates  \cite{Clerckx_RS2}. Given the channel estimate $\hat{\bs{h}}_k$ and the known error variance $\sigma_{e,k}^2$, the BS generates $N_s$ samples of the channel such that the $n$th sample is given by
\begin{equation*}
    \bs{h}_k^{(n)}=\sqrt{1-\sigma_{e,k}^2}\hat{\bs{h}}_k + \sqrt{\sigma_{e,k}^2} \tilde{\bs{h}}_k 
\end{equation*}
where $\tilde{\bs{h}}_k \sim \mathcal{N}_\mathbb{C}(\bs{0}, \bs{I}_M)$.\\
The average common and private rates are then approximated using these samples according to $\bar{R}_{\tc,k}^{(N_s)}=\frac{1}{N_s}\sum_{n=1}^{N_s}  R_{\tc,k}^\text{inst}(\bs{H}^{(n)})$ and $\bar{R}_{\tc,k}^{(N_s)}=\frac{1}{N_s}\sum_{n=1}^{N_s}  R_{\tp,k}^\text{inst}(\bs{H}^{(n)})$, respectively, with $\bs{H}^{(n)}=[\bs{h}_1^{(n)},\dots,\bs{h}_K^{(n)}]$.\\
Similar to \eqref{eq:XiRate}, a relationship between the approximated rates and the approximated augmented weighted MMSEs can be established
\begin{equation}
    \bar{\xi}_{\tc,k}^{\text{MMSE}(N_s)}=1-\bar{R}_{\tc,k}^{(N_s)}, \quad \bar{\xi}_{\tp,k}^{\text{MMSE}(N_s)}=1-\bar{R}_{\tp,k}^{(N_s)}
\end{equation}
where 
\begin{align*}
&\bar{\xi}_{\tc,k}^{\text{MMSE}(N_s)}=\frac{1}{N_s} \sum_{n=1}^{N_s} \xi_{\tc,k}^{\text{MMSE}}(\bs{H}^{(n)})\\
&\bar{\xi}_{\tp,k}^{\text{MMSE}(N_s)}=\frac{1}{N_s} \sum_{n=1}^{N_s} \xi_{\tp,k}^{\text{MMSE}}(\bs{H}^{(n)}).
\end{align*}
We can hence formulate the optimization problem based on the approximated augmented weighted MSEs
\begin{equation}
    \begin{aligned}
    \underset{\substack{\bar{\xi}_\tc, \bs{U}, \bs{G}\\\bs{p}_\tc, \bs{p}_{\tp,1},\dots, \bs{p}_{\tp,K}}}{\min} \: & \bar{\xi}_\tc + \sum_{k=1}^K \bar{\xi}_{\tp,k}^{(N_s)}\\
  \text{s.t.} \quad  &\bar{\xi}_\tc\geq \bar{\xi}_{\tc,k}^{(N_s)}, \forall k\\
    &\|\bs{p}_\tc\|^2+\sum_{k=1}^K \|\bs{p}_{\tp,k}\|^2 \leq P_\text{dl}
    \end{aligned}
    \label{eq:OptXi}
\end{equation}
where the weights and equalizing filters corresponding to the $N_s$ samples are collected respectively in $\bs{U}$ and $\bs{G}$.
\par The optimization problem \eqref{eq:OptXi} is convex in each of its blocks ($\bs{U}$, $\bs{G}$, and $\bs{P}=[\bs{p}_\tc, \bs{p}_{\tp,1}, \dots, \bs{p}_{\tp,K}]$) if the other two are fixed. Therefore, an alternating solution approach is used. For fixed precoders, $\bs{U}$ and $\bs{G}$ are computed according to the MMSE solution as explained in the beginning of this section. The weights and equalizing filters are calculated for each of the $N_s$ samples and subsequently averaged out. Then for fixed $\bs{U}$ and $\bs{G}$, the optimization problem with respect to $\bs{P}$ reduces to a quadratically constrained convex problem, which can be solved using standard interior-point solvers (e.g. CVX). We refer the interested reader to \cite{Clerckx_RS2} for a detailed description of this approach.

\section{Numerical Results}
\label{sec5}
We consider the 3GPP spatial channel model and specifically the urban microcell environment \cite{ETSI}. The BS is equipped with $M=64$ antennas and serves $K=5$ users distributed inside the cell with radius $250~\text{m}$. The users' positions are generated uniformly at random inside the cell, around the BS located at the cell center. The channels are non-line-of-sight (NLOS) and sampled from the Gaussian distribution with zero mean and covariance matrix generated according to the multi-path channel model. Assuming $N_\text{path}=6$ main clusters and $N_\text{rays}=20$ rays in each cluster, the covariance matrix of user $k$ at carrier frequency $f$ is given by
\begin{equation*}
    \boldsymbol{C}_k=p_k \sum_{n=1}^{N_\text{paths}} \frac{\beta_n}{N_\text{rays}}\sum_{m=1}^{N_\text{rays}} \boldsymbol{a}(\theta_{k,n,m},f)\boldsymbol{a}^\HH(\theta_{k,n,m},f)
\end{equation*}
where $p_k$ depends on the path-loss and, therefore, on the distance between the BS and the mobile station (MS) and $\beta_n$ denotes the power of the $n$th cluster.\\
$\boldsymbol{a}(\theta_{k,n,m},f)$ is the array steering vector. We assume that the BS antennas are in a uniform linear array (ULA) configuration and that the antenna spacing is given by half the wavelength of the UL channel. The ratio of the DL and UL carrier frequencies is denoted by $\nu=\frac{f_\text{dl}}{f_\text{ul}}=1.1$. 
The $i$th component of the array steering vector evaluated at the DL carrier frequency is given by 
\begin{align*}
    [\boldsymbol{a}^\text{dl}(\theta_{k,n,m})]_i&=[\boldsymbol{a}(\theta_{k,n,m},f_\text{dl})]_i\\ &=\exp\left(\mathrm{j}\pi \nu (i-1)\sin(\theta_{k,n,m})\right)
\end{align*}
where the angle $\theta_{k,n,m}$ is is the angle-of-arrival (AOA) of the $m$th ray inside the $m$th cluster for the $k$th user.

We assume that $T_\text{dl}=8$ out of $T_\text{coh}=200$ channel accesses in each channel coherence interval are used for the DL training and that the noise variance of the training signal is given by $\sigma_n^2=\frac{1}{P_\text{tr,dl}}$. Here, $P_\text{tr,dl}$ denotes the DL training power given by $P_\text{dl}T_\text{dl}$.
\par In the first part of this section, we assume that the BS disposes perfect knowledge of the channel covariance matrices. In the latter part, we will assess the performance obtained when only an estimate of the second-order information is available at the BS.
\par For the private precoder optimization using the proposed iterative solution approach, a maximum of $20$ iterations is used, whereas for the common precoder optimization $30$ iterations are considered at maximum. The common and private transformation matrices are initialized using the training matrix $\bs{\Phi}$ such that the power constraints are fulfilled
\begin{equation}
    \begin{aligned}
    &\bs{a}_{\tc,k}=\text{vec}(\bs{A}_{\tc,k})=\text{vec}(\bs{\Phi}),  \quad &&\bs{a}_{\tp,k}=\text{vec}(\bs{A}_{\tp,k})=\text{vec}(\bs{\Phi}) \\
    &  \bs{a}_\tc'=[\bs{a}_{\tc,1}^\TT,\dots,\bs{a}_{\tc,K}^\TT]^\TT, \quad &&\bs{a}_\tp'=[\bs{a}_{\tp,1}^\TT,\dots,\bs{a}_{\tp,K}^\TT]^\TT\\
    &\bs{a}_\tc= \frac{\sqrt{\alpha_\tc P_{\text{dl}}}}{\|\bs{a}_\tc'\|_2} \bs{a}_\tc', \quad &&\bs{a}_\tp= \frac{\sqrt{(1-\alpha_\tc) P_{\text{dl}}}}{\|\bs{a}_\tp'\|_2} \bs{a}_\tp'
\end{aligned}
\end{equation}
\par The results are presented in terms of the average achievable sum rate in bits per channel use as function of the DL transmit power. The expressions for the average sum rate achieved with and without applying the RS approach are obtained as follows
		\begin{equation}
		    \label{eq:RsumRSnoRS}
		    \begin{aligned}
			&R_\text{sum}^\text{RS}=\tau \left(\mathrm{E}[\log_2\left(1+\gamma_\text{c}^\text{inst}\right)]+\sum_{k=1}^K \mathrm{E}[ \log_2\left(1+\gamma^\text{inst}_{\text{p},k}\right) ] \right)\\
			&R_\text{sum}^\text{noRS}=\tau \sum_{k=1}^K\mathrm{E}[ \log_2\left(1+\gamma_{\text{p},k}^\text{inst}\right)]
		\end{aligned}
		\end{equation}
		where the instantaneous common and private SINRs are respectively given by
		\begin{equation}
		    \label{eq:InstSINRs}
		    \begin{aligned}
			&\gamma^\text{inst}_{\text{c},k}=\frac{|\boldsymbol{h}_k^\mathrm{H}\boldsymbol{p}_\text{c} |^2}{\sum_{j=1}^K |\boldsymbol{h}_k^\mathrm{H}\boldsymbol{p}_{\text{p},j}|^2+1},\\
			&\gamma_{\text{p},k}^\text{inst}=\frac{|\boldsymbol{h}_k^\mathrm{H}\boldsymbol{p}_{\text{p},k} |^2}{\sum_{j\neq k} |\boldsymbol{h}_k^\mathrm{H}\boldsymbol{p}_{\text{p},j}|^2+1}
		\end{aligned}
		\end{equation}
and $\gamma_\text{c}^\text{inst}=\underset{k}{\min}\: \gamma_{\text{c},k}^\text{inst}$. \\
The pre-log factor  $\tau=\left(1-\frac{T_\text{dl}}{T_\text{coh}}\right)$ in \eqref{eq:RsumRSnoRS} accounts for DL training overhead resulting from dedicating
$T_\text{dl}$ out of $T_\text{coh}$ channel uses in each channel coherence block for sending the pilot sequences. The overhead resulting from the CSI feedback has to be accounted for, when considering the UL setup.\\
The expectations in \eqref{eq:RsumRSnoRS} are evaluated through Monte Carlo simulations over $N_\text{iter}=300$~channel realizations.

First, we compare the results for the methods used for the common precoder optimization, which we presented in Section \ref{sec:ComPreOpt} to solve \eqref{eq:ComOpt3} for fixed auxiliary variable $\eta_k$. The first approach is to use the CVX-solver, since \eqref{eq:ComOpt3} is convex in $\bs{a}_\tc$ for fixed $\eta_k$. The second approach follows the steps described in Algorithm~\ref{alg:IncSINR2}. In Fig.~\ref{fig:ComCompare}, one can observe that there is barely any difference in terms of the achievable sum rate between both approaches. However, as depicted in Fig.~\ref{fig:ConvergenceComp}, where the minimum common SINR is plotted versus the iteration number, the iterative increasing SINR method converges faster than the approach based on using the CVX-solver and will therefore be used for the remaining part of the presented results. 
\begin{figure}[h] 
\centering
	\scalebox{0.5}{\input{Common_Comp_SumRate}}
	\caption{Comparison of the proposed methods for common precoder optimization in terms of the sum rate versus the DL power}
	\label{fig:ComCompare}
\end{figure}
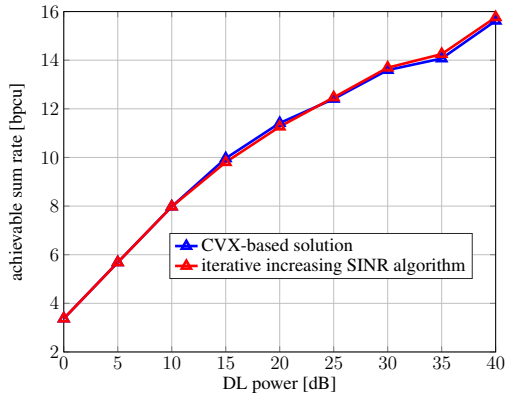
\begin{figure}[h]
\centering
	\scalebox{0.5}{\input{Convergence_Objective}}
	\caption{Convergence behaviour of the proposed methods for common precoder optimization}
	\label{fig:ConvergenceComp}
\end{figure}
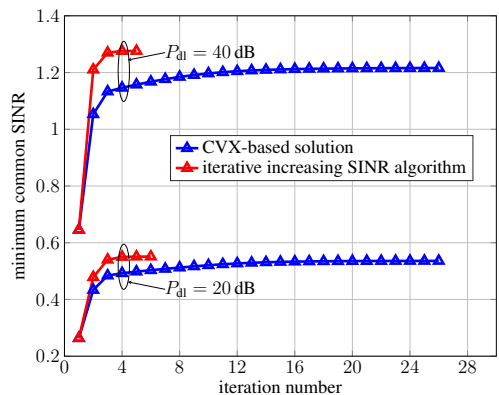

In Fig.~\ref{fig:SR_LB_RS}, we depict the average achievable sum rate as well as the lower bound (LB) on the achievable rate which we used as a figure-of-merit in our derivations
\begin{align*}
			&R_\text{sum}^\text{LB,RS}=
			\tau \left(\log_2\left(1+\gamma_\text{c}\right)+\sum_{k=1}^K \log_2\left(1+\gamma_{\text{p},k}\right) \right)\\
			&R_\text{sum}^\text{LB,noRS}=\tau \sum_{k=1}^K \log_2\left(1+\gamma_{\text{p},k}\right)
	\end{align*}
where $\gamma_\tc=\underset{k}{\min}\gamma_{\tc,k}$ and the SINRs $\gamma_{\tc,k}$ and $\gamma_{\text{p},k}$ are evaluated according to \eqref{eq:SINRc} and \eqref{eq:SINRp}, respectively.\\
One can easily observe that the RS approach is not only beneficial for the LB but also more significantly for the actual achievable sum rate. While the achievable sum rate obtained without applying RS saturates in the high power regime at approximately $11.5$~dB, RS clearly leads to an increase of the rate as the transmit power increases. When the system is interference-limited in the high power regime, the sum rate can be increased by investing more power into transmitting the common message. The gains leveraged by the RS approach are visible in the medium power range and more prominent in the high power regime. For $P_\text{dl}=40$~dB, the RS scheme leads to an increase in the sum rate of more than $4$~bits per channel use (bpcu) compared to the non-RS scheme. 
\begin{figure}[h]
	\centering
	\scalebox{0.5}{\input{Bilin_Achievable_Rate_vs_LB_RS_noRS}}
	\caption{Average achievable sum rate and lower bound on the achievable rate for the bilinear precoding approach w/ and w/o RS}
	\label{fig:SR_LB_RS}
\end{figure}
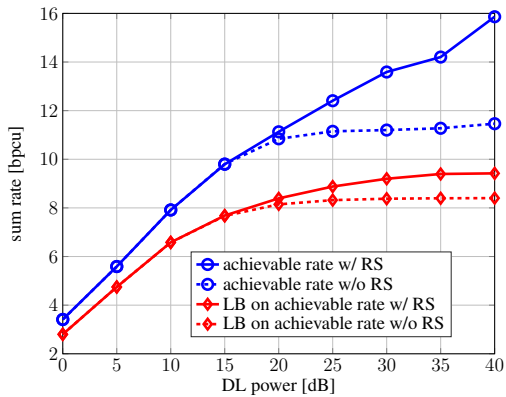
\begin{figure}[h]
\centering
	\scalebox{0.5}{\input{UE_private_common_rates_Bilin}}
	\caption{Illustration of the individual private rates of the users and minimum common rate versus the downlink transmit power}
	\label{fig:UEratesCommonRate}
\end{figure}
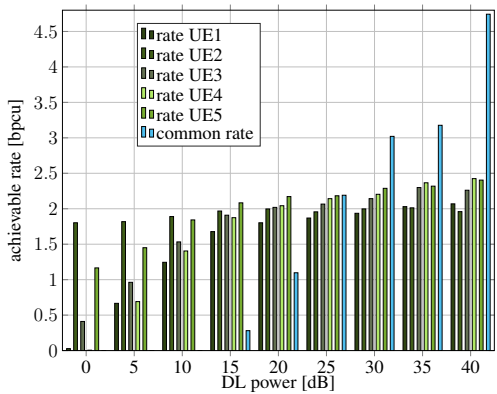
In Fig.~\ref{fig:UEratesCommonRate}, we plot the individual users' rates and the common rate, in order to visualize the effect of the common message on the overall sum rate.
We can confirm our interpretation from the previous results, by noticing that while the private rates of the users saturate starting from $P_\text{dl}=20$~dB, the common rate increases as the transmit power increases.\\

\begin{figure}[h]
\centering
	\scalebox{0.5}{\input{Results_K=2_K=8}}
	\caption{Average achievable sum rate for a system with $M=32$ BS antennas and $T_\text{dl}=8$ DL pilots with two different numbers of users $K=2$ and $K=8$ where the bilinear precoding approach w/ or w/o RS is applied}
	\label{fig:RatewrtK}
\end{figure}
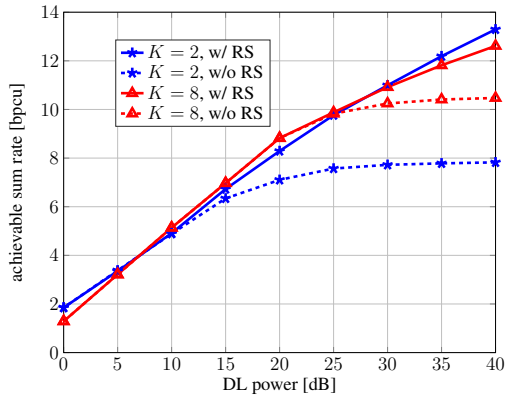
Next, we consider the system with $M=32$ BS antennas, $T_\text{dl}=8$ DL pilots and different numbers of users. In Fig.~\ref{fig:RatewrtK}, we plot the average achievable sum rate of the system with and without RS for $K=2$ and $K=8$. One could observe the advantage of RS decreases drastically as the number of users increases, since the common stream has to be decoded by all users and therefore the minimum common rate decreases.

\begin{figure}[h]
\centering
	\scalebox{0.5}{\input{Results_M=32_M=16}}
	\caption{Average achievable sum rate for a system with $K=4$ users and different numbers of BS antennas $M=32$ and $M=16$ and $T_\text{dl}=M/4$ DL pilots where the bilinear precoding approach w/ or w/o RS is applied}
	\label{fig:RatewrtM}
\end{figure}
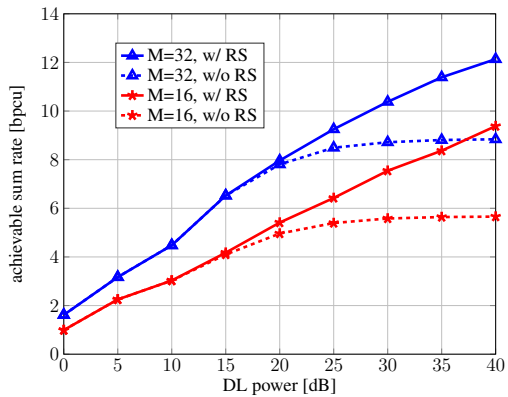

In Fig.~\ref{fig:RatewrtM}, we additionally plot the achievable average sum rate for two different numbers of BS antennas, namely $M=16$ and $M=32$. The number of DL pilots $T_\text{dl}$ is set to $M/4$ in both cases while the number of users for both setups is $K=4$. One can observe as shown in the previous results that in the two cases, the RS enhances the system performance at medium to high power values. The system with less antennas ($M=16$) seems to benefit from the application of RS already at transmit powers less than $20$~dB, whereas the benefits of RS for the case $M=32$ become more visible at transmit powers higher than $20$~dB.
\subsection{Comparison to the IWMMSE Approach}
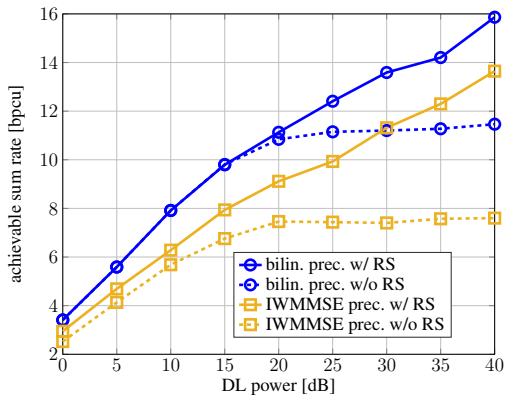
\begin{figure}[h]
\centering
	\scalebox{0.5}{\input{Bilin_IWMMSE_RS_vs_noRS}}
	\caption{Comparison of the average achievable sum rate obtained when applying the bilinear precoding against the IWMMSE approach w/ and w/o RS}
	\label{fig:IWMMSE_BILIN}
\end{figure}

For the IWMMSE approach, we use the MMSE channel estimate $\hat{\bs{h}}_k^\text{MMSE}$ that we obtain from the channel observation $\bs{y}_k$ according to
\begin{align*}
    \hat{\bs{h}}_k^\text{MMSE}= \bs{C}_k \bs{\Phi} (\bs{\Phi}^\HH \bs{C}_k \bs{\Phi} + \sigma_n^2 \bs{I}_{T_\text{dl}})^{-1} \bs{y}_k.
\end{align*}
In order to generate the $N_s=200$~samples used for SAA, the estimation error variance is calculated based on the channel second-order statistics $\sigma_{e,k}^2=\text{tr}(\bs{C}_k-\bs{C}_k \bs{\Phi} (\bs{\Phi}^\HH \bs{C}_k \bs{\Phi} + \sigma_n^2 \bs{I}_{T_\text{dl}})^{-1} \bs{\Phi}^\HH \bs{C}_k)/M$. A maximum of $30$~iterations is used for the alternating approach. As initialization for the IWMMSE approach, we use the MMSE channel estimate for the private precoders. As for the common precoder, we use the dominant left singular vector of the MMSE estimate of the channel matrix. The power fraction obtained using the bilinear approach is taken as initial value for the IWMMSE, where uniform power allocation is performed among the private streams.\\ 
In Fig.~\ref{fig:IWMMSE_BILIN}, we depict the results in terms of the average achievable sum rate as a function of the DL transmit power for both cases, when the RS scheme is applied and when it is not.  \\
\begin{figure}
\centering
	\scalebox{0.5}{\input{Bilin_IWMMSE_Known_vs_Estimated_Cov_Matrices}}
	\caption{Comparison of the average achievable sum rate obtained for perfectly known versus estimated channel covariance matrices w/ RS}
	\label{fig:KnownEst}
\end{figure}
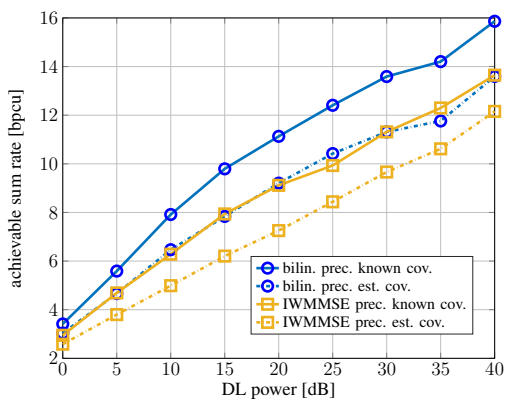
Obviously, the bilinear precoding approach outperforms the IWMMSE method in both cases, i.e., with or without RS. Both methods seem to benefit from RS in the medium to high power regime. While for the bilinear approach the gains are visible starting from $P_\text{dl}=20$~dB, the RS scheme significantly improves the rate of the IWMMSE approach already for $P_\text{dl}=15$~dB. When only private messages are sent to the users, both rates saturate in the high power regime as observed before and the gap remains almost constant. \\
In the high power regime, the bilinear approach combined with RS leads to a slope of about $0.33$~bpcu/dB, which is slightly higher than the one for the IWMMSE approach given by approximately $0.27$~bpcu/dB. \\
We need to mention here, that the optimization problem for the bilinear precoding approach considers a relaxed power constraint, whereas the power constraint for the IWMMSE approach is an instantaneous one. This offers more degrees of freedom for the bilinear precoder design. However, even after scaling the bilinear precoders to satisfy the instantaneous total power constraint, the change in terms of the achievable sum rate was negligible. 
\par Regarding the runtimes of both algorithms, the overall optimization of the transformation matrices of the bilinear precoders converges after $1431$~s \textbf{for a given realisation of the channel covariance matrices} at a DL transmit power of $P_\text{dl}=40$~dB. On the other hand, the runtime of the IWMMSE approach \textbf{for a given channel realisation} is about $184$~s at $P_\text{dl}=40$~dB. Obviously, the runtime needed for the bilinear approach is much larger than the one required for the IWMMSE method. However,
we need to emphasize, that the optimization problem proposed for finding the deterministic transformations defining the bilinear precoder has only to be solved once the covariance matrices change and therefore only a matrix-vector multiplication is performed to find the precoding vector for each channel realization. Contrary, the IWMMSE method has to be run in each channel coherence interval. It is also noteworthy that the IWMMSE approach is very sensitive to the initialization of the precoders especially in the high power regime. 
\subsection{Comparison of the Performance for Perfectly Known vs. Estimated Covariance Matrices}
The performance of the system with perfect knowledge of the covariance matrices shall now be compared to the one when the BS disposes not the true DL channel covariance matrices but an estimate thereof. For the estimation procedure, we use the extrapolation approach proposed in our previous work \cite{Donia_WSA20}. In a few words, this approach consists of exploiting a reciprocity relationship between the UL and DL covariance matrices in order to estimate the DL covariance matrices using the UL channel observations.
From Fig.~\ref{fig:KnownEst}, one can see that when estimated covariance matrices are used, a performance degradation of about $1$~bpcu in the low power regime is experienced and the gap becomes about $2$~bpcu in the medium to high power range. \\
Nevertheless, we observe that the slope of the curves remains the same for high power values, which confirms the robustness of the applied covariance estimation approach.

\section{Conclusion}
In this work, we presented a framework for the application of rate splitting to an FDD massive multi-user MIMO system. Despite the suboptimality of the proposed approach and the incomplete channel knowledge at the BS, numerical results have shown that the combination of the bilinear precoding based on the second-order channel information and rate splitting leads to a drastic improvement in terms of the achievable sum rate when compared to the system without rate splitting. The transmission of the common message is mainly beneficial when it comes to medium to high transmit powers. While the conventional DL system (usual broadcast channel, without rate splitting) saturates in the high transmit power regime due to the inter-user interference and the imperfect channel knowledge, introducing the common stream yields a rate of the system that increases when more transmit power is used.  Furthermore, our method significantly outperforms the commonly used iterative weighted MMSE approach. While the optimization problem related to the latter has to be solved in each and every channel coherence interval, the bilinear approach depending on the channel covariance matrices requires a simple matrix-vector multiplication in every channel coherence interval. Finding the deterministic transformation matrices that constitute the bilinear precoder has only to be performed in the beginning of the covariance coherence interval which is given by several channel coherence intervals.\\
A future research direction would be to extend the proposed approach to a multi-cell setup where the challenges related to the system design become more significant due to the inter-cell interference. Furthermore, since the benefits of RS decrease as the number of served users increases, one could optimize the set of users that have to decode the common stream, which could increase the minimum common rate. This user selection can be based on the channel second-order statistics.

\appendices
%\newpage
% you can choose not to have a title for an appendix
% if you want by leaving the argument blank
\section{Derivation of the Closed-Form SINR expressions for the Bilinear Approach} \label{app1}
\input{Appendix1}

\section{Derivation of the Parameters of the SINR Increasing Approach} \label{app2}
\input{Appendix2}

%The authors would like to thank...

% Can use something like this to put references on a page
% by themselves when using endfloat and the captionsoff option.
\ifCLASSOPTIONcaptionsoff
  \newpage
\fi

\end{document}

%% file: SystemModelTx.tex
\tikzstyle{int}=[draw, rectangle,minimum width=1cm, minimum height=2.5cm, thick]
\tikzstyle{init} = [pin edge={<-,thick,black}]
\tikzstyle{sum} = [draw, circle,inner sep=1pt, minimum size=2mm, very thick] % Adder

\begin{tikzpicture}[node distance=2cm,auto,>=latex']
    \node [int] (s) [minimum height=5cm] {Split};
    \node [int] (c) at ([xshift=1.5cm, yshift=1.5cm]s.east) {Combine};
    \node [int] (e) [minimum height=5cm] at ([xshift=1.5cm,yshift=-1.5cm]c.east) {Encode};
    \node [int] (p) [minimum height=5cm] at ([xshift=1.5cm]e.east) {Precode};
    
    \node (dots) at ([xshift=-0.5cm, yshift=0.2cm]s.west) {$\vdots$};
    
     \draw[-latex,thick] ([xshift=-0.75cm,yshift=1cm]s.west) to node[midway,above=0.01cm]{$W_1$} ([yshift=1cm]s.west);
    \draw[-latex,thick] ([xshift=-0.75cm,yshift=-1cm]s.west) to node[midway,above=0.01cm]{$W_K$} ([yshift=-1cm]s.west);
    
      \node (dots) at ([xshift=-0.35cm, yshift=0.1cm]c.west) {$\vdots$};
    
     \draw[-latex,thick] ([yshift=2cm]s.east) to node[midway,above=0.01cm]{$W_{\tc,1}$} ([yshift=0.5cm]c.west);
     \draw[-latex,thick] ([yshift=0.5cm]s.east) to node[midway,above=0.01cm]{$W_{\tc,K}$} ([yshift=-1cm]c.west);
     
      \draw[-latex,thick] ([yshift=-0.5cm]s.east) to node[midway,above=0.01cm]{$W_{\tp,1}$} ([yshift=-0.5cm]e.west);
     \draw[-latex,thick] ([yshift=-2cm]s.east) to node[midway,above=0.01cm]{$W_{\tp,K}$} ([yshift=-2cm]e.west);
     
     \node (dots) at ([xshift=1.5cm, yshift=-1cm]s.east) {$\vdots$};
     
      \draw[-latex,thick] (c.east) to node[midway,above=0.01cm]{$W_{\tc}$} ([yshift=1.5cm]e.west);
      
       \draw[-latex,thick] ([yshift=1.5cm]e.east) to node[midway,above=0.01cm]{$s_{\tc}$} ([yshift=1.5cm]p.west);
       
        \draw[-latex,thick] ([yshift=-0.5cm]e.east) to node[midway,above=0.01cm]{$s_{\tp,1}$} ([yshift=-0.5cm]p.west);
     \draw[-latex,thick] ([yshift=-2cm]e.east) to node[midway,above=0.01cm]{$s_{\tp,K}$} ([yshift=-2cm]p.west);
     
     \node (dots) at ([xshift=0.35cm,yshift=-1cm]e.east) {$\vdots$};

      \draw[-latex,thick] (p.east) to node[midway,above=0.01cm]{$\bs{x}$} ([xshift=0.75cm]p.east);

\end{tikzpicture}

%% file: SystemModelMS.tex
\tikzstyle{int}=[draw, rectangle,minimum width=1cm, minimum height=1.5cm, thick, text width=1.5cm]
\tikzstyle{init} = [pin edge={<-,thick,black}]
\tikzstyle{sum} = [draw, circle,inner sep=1pt, minimum size=2mm, very thick] % Adder

\begin{tikzpicture}[node distance=2cm,auto,>=latex']
    \node [int] (e1) [align=center,minimum height=1.5cm] {Equalize + Decode};
    \node [int] (s) [align=center] at ([xshift=1cm, yshift=-2.5cm]e1.east) {SIC};
    \node [int] (c) [align=center,minimum height=1.5cm] at ([xshift=4cm]e1.east) {Combine};
     \node [int] (e2) [align=center,minimum height=1.5cm] at ([yshift=-2.5cm]c) {Equalize + Decode};

     \draw[-latex,thick] ([xshift=-0.75cm]e1.west) to node[midway,above=0.01cm]{$r_{\tc,k}$} (e1.west);
   
    \draw[-latex,thick] (e1.east) to node[midway,above=0.01cm]{$\hat{W}_{\tc}$}  ([xshift=1.5cm]e1.east) to (c.west);
    
    \draw[-latex,thick] ([xshift=-0.35cm]e1.west) to ([xshift=-0.35cm,yshift=-2.5cm]e1.west) to (s.west);
    
    \draw[-latex,thick] (c.east) to node[midway,above=0.01cm]{$\hat{W}_{k}$} ([xshift=0.75cm]c.east);
 
     \draw[-latex,thick] ([xshift=1cm]e1.east) to (s.north);
     
     \draw[-latex,thick] (s.east) to node[midway,above=0.01cm]{$r_{\tp,k}$} (e2.west);
     
      \draw[-latex,thick] (e2.north) to node[midway,right=0.01cm]{$\hat{W}_{\tp,k}$}  (c.south);

\end{tikzpicture}

%% file: Common_Comp_SumRate.tex
% This file was created by matlab2tikz.
%
%The latest updates can be retrieved from
%  http://www.mathworks.com/matlabcentral/fileexchange/22022-matlab2tikz-matlab2tikz
%where you can also make suggestions and rate matlab2tikz.
\definecolor{mycolor1}{rgb}{0.00000,0.44700,0.74100}%
\definecolor{mycolor2}{rgb}{0.85000,0.32500,0.09800}%
\definecolor{mycolor3}{rgb}{0.92900,0.69400,0.12500}%
\definecolor{mycolor4}{rgb}{0.49400,0.18400,0.55600}%
\begin{tikzpicture}

\begin{axis}[%
width=4.521in,
height=3.566in,
at={(0.758in,0.481in)},
scale only axis,
xmin=0,
xmax=40,
xticklabel style={font=\Large},
yticklabel style={font=\Large},
xlabel style={font=\color{white!15!black},font=\Large},
xlabel={DL power [dB]},
ymin=2,
ymax=16,
ylabel style={font=\color{white!15!black},font=\Large},
ylabel={achievable sum rate [bpcu]},
axis background/.style={fill=white},
xmajorgrids,
ymajorgrids,
legend style={at={(0.245,0.215)}, anchor=south west, legend cell align=left, align=left, draw=white!15!black,font=\Large}
]
\addplot [color=blue, line width=2pt, mark=triangle, mark options={solid, blue}, mark size=4pt]
  table[row sep=crcr]{%
0	3.37293098872696\\
5	5.68384136201219\\
10	7.97758561722117\\
15	9.96326455508101\\
20	11.4148129073649\\
25	12.4071859207363\\
30	13.5939511308361\\
35	14.0692359903316\\
40	15.6206189347663\\
};
\addlegendentry{CVX-based solution}

\addplot [color=red, line width=2pt,mark=triangle, mark options={solid, red}, mark size=4pt]
  table[row sep=crcr]{%
0	3.37182204979664\\
5	5.68362007042906\\
10	7.97714226102312\\
15	9.80498451374552\\
20	11.2662184276306\\
25	12.4681563630372\\
30	13.6920422952164\\
35	14.2505458889641\\
40	15.7635675472856\\
};
\addlegendentry{iterative increasing SINR algorithm}

\end{axis}

\end{tikzpicture}%

%% file: Convergence_Objective.tex
% This file was created by matlab2tikz.
%
%The latest updates can be retrieved from
%  http://www.mathworks.com/matlabcentral/fileexchange/22022-matlab2tikz-matlab2tikz
%where you can also make suggestions and rate matlab2tikz.
%
\definecolor{mycolor1}{rgb}{0.00000,0.44700,0.74100}%
\definecolor{mycolor2}{rgb}{0.85000,0.32500,0.09800}%
\definecolor{mycolor3}{rgb}{0.00000,0.44706,0.74118}%
\definecolor{mycolor4}{rgb}{0.85098,0.32549,0.09804}%
\begin{tikzpicture}

\begin{axis}[%
width=4.521in,
height=3.566in,
at={(0.758in,0.481in)},
scale only axis,
xmin=0,
xmax=30,
ymin=0.2,
ymax=1.4,
xtick={0,4,8,12,16,20,24,28},
xticklabel style={font=\Large},
yticklabel style={font=\Large},
xlabel style={font=\color{white!15!black},font=\Large},
xlabel={iteration number},
ylabel style={font=\color{white!15!black},font=\Large},
ylabel={minimum common SINR},
xmajorgrids,
ymajorgrids,
axis background/.style={fill=white},
legend style={at={(0.245,0.515)}, anchor=south west, legend cell align=left, align=left, draw=white!15!black,font=\Large}
]
\addplot [color=blue, line width=2pt,mark=triangle, mark options={solid, blue}, mark size=4pt]
  table[row sep=crcr]{%
1	0.645710800494879\\
2	1.05289955501983\\
3	1.13307706425421\\
4	1.1461319032534\\
5	1.15734503299286\\
6	1.16757500343494\\
7	1.17665491589631\\
8	1.18459682053716\\
9	1.19129561140709\\
10	1.196778794215\\
11	1.20116153393728\\
12	1.20460260107718\\
13	1.20727061857993\\
14	1.20932300070505\\
15	1.21089568399045\\
16	1.21210016463573\\
17	1.21302477975598\\
18	1.21373784519254\\
19	1.21429133589604\\
20	1.21472436762947\\
21	1.2150661914091\\
22	1.21533860835922\\
23	1.21555785816427\\
24	1.21573605952947\\
25	1.21588227643973\\
26	1.2160033306292\\
};
\addlegendentry{CVX-based solution}

\addplot [color=red, line width=2pt,mark=triangle, mark options={solid, red}, mark size=4pt]
  table[row sep=crcr]{%
1	0.645710800494877\\
2	1.21016111297297\\
3	1.26985625887322\\
4	1.27593242128698\\
5	1.27606085621978\\
};
\addlegendentry{iterative increasing SINR algorithm}

\addplot [color=blue, line width=2pt,mark=triangle, mark options={solid, blue}, mark size=4pt]
 table[row sep=crcr]{%
1	0.26350924302873\\
2	0.433570956514143\\
3	0.484241745183187\\
4	0.49255292920261\\
5	0.498189919342211\\
6	0.503145412433762\\
7	0.507810970791616\\
8	0.512378048606542\\
9	0.51672206549254\\
10	0.520685354637561\\
11	0.524136420192717\\
12	0.527008660186047\\
13	0.529305793004292\\
14	0.531084012962917\\
15	0.53242630434555\\
16	0.533421050938488\\
17	0.534148946523971\\
18	0.534677298961446\\
19	0.535059111710778\\
20	0.535334578286985\\
21	0.535533424552657\\
22	0.535677279482568\\
23	0.535781717206707\\
24	0.535857882624112\\
25	0.535913725953668\\
26	0.535954913130785\\
};
%\addlegendentry{data3}

\addplot [color=red, line width=2pt,mark=triangle, mark options={solid, red}, mark size=4pt]
  table[row sep=crcr]{%
1	0.26350924302873\\
2	0.478508182325572\\
3	0.540248915878086\\
4	0.549928645101518\\
5	0.551062200633798\\
6	0.551066355247453\\
};
%\addlegendentry{data4}

\end{axis}

\draw [black] (3.5,8.8) ellipse [x radius=0.15, y radius=0.8];
\draw [black] (3.5,3.6) ellipse [x radius=0.15, y radius=0.6];
\draw [-stealth](4.6,9.3) -- (3.6,9.1);
\node at (5.8,9.3) {\Large $P_\text{dl}=40\,\text{dB}$};
\draw [-stealth](4.6,3) -- (3.6,3.2);
\node at (5.8,3) {\Large $P_\text{dl}=20\,\text{dB}$};
\end{tikzpicture}%

%% file: Bilin_Achievable_Rate_vs_LB_RS_noRS.tex
% This file was created by matlab2tikz.
%
%The latest updates can be retrieved from
%  http://www.mathworks.com/matlabcentral/fileexchange/22022-matlab2tikz-matlab2tikz
%where you can also make suggestions and rate matlab2tikz.
%
\definecolor{mycolor1}{rgb}{0.00000,0.44700,0.74100}%
\definecolor{mycolor2}{rgb}{0.00000,0.44706,0.74118}%
\definecolor{mycolor3}{rgb}{0.00000,0.44706,0.74118}%
\definecolor{mycolor4}{rgb}{0.85098,0.32549,0.09804}%
\begin{tikzpicture}

\begin{axis}[%
width=4.521in,
height=3.566in,
at={(0.758in,0.481in)},
scale only axis,
xmajorgrids,
ymajorgrids,
xmin=0,
xmax=40,
ymin=2,
ymax=16,
xticklabel style={font=\Large},
yticklabel style={font=\Large},
xlabel style={font=\color{white!15!black},font=\Large},
ylabel style={font=\color{white!15!black},font=\Large},
xlabel={DL power [dB]},
ylabel={sum rate [bpcu]},
axis background/.style={fill=white},
legend style={at={(0.9,0.3)},legend cell align=left, align=left, draw=white!15!black,font=\Large}
]
\addplot [color=blue, line width=2pt,mark=o, mark options={solid, blue}, mark size=4pt]
  table[row sep=crcr]{%
0	3.41316990712724\\
5	5.58830868705795\\
10	7.91493546148591\\
15	9.79351608509658\\
20	11.1303474838492\\
25	12.407690445371\\
30	13.589273452135\\
35	14.2038859855199\\
40	15.8623926005369\\
};
\addlegendentry{achievable rate w/ RS}

\addplot [color=blue, dashed, line width=2pt,mark=o, mark options={solid, blue}, mark size=4pt]
  table[row sep=crcr]{%
0	3.40792413844249\\
5	5.58899348947436\\
10	7.91514675823873\\
15	9.80426952886771\\
20	10.8414409091281\\
25	11.1496506081529\\
30	11.2013829813646\\
35	11.2752788288583\\
40	11.462833952141\\
};
\addlegendentry{achievable rate w/o RS}

\addplot [color=red, line width=2pt ,mark=diamond, mark options={solid, red}, mark size=4pt]
  table[row sep=crcr]{%
0	2.80030179785502\\
5	4.74674429001292\\
10	6.58144153792129\\
15	7.68518138647206\\
20	8.39234050271777\\
25	8.87846990059559\\
30	9.1985134523751\\
35	9.39507848704996\\
40	9.42048617057549\\
};
\addlegendentry{LB on achievable rate w/ RS}

\addplot [color=red, dashed, line width=2pt, ,mark=diamond, mark options={solid, red}, mark size=4pt]
  table[row sep=crcr]{%
0	2.80061988352905\\
5	4.74689464939231\\
10	6.58159668630952\\
15	7.67005059333763\\
20	8.14590727135866\\
25	8.31959452318124\\
30	8.3776089580497\\
35	8.39630370604781\\
40	8.4022522136809\\
};
\addlegendentry{LB on achievable rate w/o RS}

\end{axis}

\end{tikzpicture}%

%% file: UE_private_common_rates_Bilin.tex
% This file was created by matlab2tikz.
%
%The latest updates can be retrieved from
%  http://www.mathworks.com/matlabcentral/fileexchange/22022-matlab2tikz-matlab2tikz
%where you can also make suggestions and rate matlab2tikz.
%
\definecolor{mycolor1}{rgb}{0.18824,0.27059,0.07843}%
\definecolor{mycolor2}{rgb}{0.25098,0.36863,0.09412}%
\definecolor{mycolor3}{rgb}{0.38824,0.43922,0.32157}%
\definecolor{mycolor4}{rgb}{0.69020,0.90980,0.40000}%
\definecolor{mycolor5}{rgb}{0.46600,0.67400,0.18800}%
\definecolor{mycolor6}{rgb}{0.30100,0.74500,0.93300}%

\begin{tikzpicture}

\begin{axis}[%
width=4.521in,
height=3.566in,
at={(0.758in,0.481in)},
scale only axis,
ybar=2.5pt,
bar width=0.1cm,
xmin=0.506666666666667,
xmax=9.49333333333333,
xtick={1,2,3,4,5,6,7,8,9},
xticklabels={{0},{5},{10},{15},{20},{25},{30},{35},{40}},
xticklabel style={font=\Large},
yticklabel style={font=\Large},
xlabel style={font=\color{white!15!black},font=\Large},
ylabel style={font=\color{white!15!black},font=\Large},
xlabel={DL power [dB]},
ymin=0,
ymax=4.8,
ytick={0,0.5,1,1.5,2,2.5,3,3.5,4,4.5},
ylabel={achievable rate [bpcu]},
axis background/.style={fill=white},
xmajorgrids,
ymajorgrids,
legend style={at={(0.173,0.6)}, anchor=south west, legend cell align=left,legend columns=1, align=left, draw=white!15!black,font=\Large}
]

%\addplot[ybar, bar width=0.107, fill=mycolor1, draw=black, area legend] table[row sep=crcr] {%
%1	0.0276484889861395\\
%2	0.666182452827325\\
%3	1.24494751187519\\
%4	1.67783291639624\\
%5	1.80221126549281\\
%6	1.86932150448106\\
%7	1.93606389622656\\
%8	2.02984814645454\\
%9	2.06915811574924\\
%};
%\addplot[forget plot, color=white!15!black] table[row sep=crcr] {%
%0.506666666666667	0\\
%9.49333333333333	0\\
%};
\addplot[black,fill=mycolor1, draw=black] coordinates {
(1,	0.0276484889861395) (2,0.666182452827325) (3,	1.24494751187519) (4,1.67783291639624) (5,1.80221126549281) (6,1.86932150448106) (7,1.93606389622656) (8, 2.02984814645454) (9, 2.06915811574924)
  };
\addlegendentry{rate UE1}

%\addplot[ybar, bar width=0.107, fill=mycolor2, draw=black, area legend] table[row sep=crcr] {%
%1	1.8024336338559\\
%2	1.81807272808099\\
%3	1.88981467950941\\
%4	1.96795429765951\\
%5	1.9971682151301\\
%6	1.95633485196193\\
%7	1.99798492013423\\
%8	2.01293652129651\\
%9	1.95943028036962\\
%};
%\addplot[forget plot, color=white!15!black] table[row sep=crcr] {%
%0.506666666666667	0\\
%9.49333333333333	0\\
%};
\addplot[black,fill=mycolor2, draw=black] coordinates {
(1, 1.8024336338559) (2,1.81807272808099) (3,1.88981467950941) (4,1.96795429765951) (5,1.9971682151301) (6,1.95633485196193) (7,1.99798492013423) (8,2.01293652129651) (9,1.95943028036962)
  };
\addlegendentry{rate UE2}

%\addplot[ybar, bar width=0.107, fill=mycolor3, draw=black, area legend] table[row sep=crcr] {%
%1	0.410917982269085\\
%2	0.962570649096448\\
%3	1.53247179640648\\
%4	1.90921258898673\\
%5	2.01944817121375\\
%6	2.06584404256085\\
%7	2.14259006810398\\
%8	2.29914264521114\\
%9	2.26130118585628\\
%};
%\addplot[forget plot, color=white!15!black] table[row sep=crcr] {%
%0.506666666666667	0\\
%9.49333333333333	0\\
%};
\addplot[black,fill=mycolor3, draw=black] coordinates {
(1, 	0.410917982269085) (2,0.962570649096448) (3,	1.53247179640648) (4,1.90921258898673) (5,2.01944817121375) (6,2.06584404256085) (7,2.14259006810398) (8,2.29914264521114) (9, 2.26130118585628)
  };
\addlegendentry{rate UE3}

%\addplot[ybar, bar width=0.107, fill=mycolor4, draw=black, area legend] table[row sep=crcr] {%
%1	0.00539348749019891\\
%2	0.690678658112541\\
%3	1.4046406519638\\
%4	1.8740802190517\\
%5	2.04276865941613\\
%6	2.14239712560334\\
%7	2.20385782653828\\
%8	2.366205417399\\
%9	2.4249456626166\\
%};
%\addplot[forget plot, color=white!15!black] table[row sep=crcr] {%
%0.506666666666667	0\\
%9.49333333333333	0\\
%};
\addplot[black,fill=mycolor4, draw=black] coordinates {
(1, 0.00539348749019891) (2,0.690678658112541) (3,	1.4046406519638) (4,1.8740802190517) (5,2.04276865941613) (6,	2.14239712560334) (7,2.20385782653828) (8,2.366205417399) (9, 2.4249456626166)
  };
\addlegendentry{rate UE4}

%\addplot[ybar, bar width=0.107, fill=mycolor5, draw=black, area legend] table[row sep=crcr] {%
%1	1.16593816578256\\
%2	1.45063538438221\\
%3	1.84258665034906\\
%4	2.08335821371269\\
%5	2.17147125788242\\
%6	2.18318348943833\\
%7	2.28800446634776\\
%8	2.3187190827182\\
%9	2.4041787440647\\
%};
%\addplot[forget plot, color=white!15!black] table[row sep=crcr] {%
%0.506666666666667	0\\
%9.49333333333333	0\\
%};
\addplot[black,fill=mycolor5, draw=black] coordinates {
(1, 1.16593816578256) (2,1.45063538438221) (3,	1.84258665034906) (4,2.08335821371269) (5,2.17147125788242) (6,2.18318348943833) (7,2.28800446634776) (8,2.3187190827182) (9, 2.4041787440647)
  };
\addlegendentry{rate UE5}

%\addplot[ybar, bar width=0.107, fill=mycolor6, draw=black, area legend] table[row sep=crcr] {%
%1	0.000838148743364158\\
%2	0.000168814558434414\\
%3	0.000474171381969933\\
%4	0.281077849289718\\
%5	1.09727991471399\\
%6	2.19060943132545\\
%7	3.02077227478425\\
%8	3.1770341724405\\
%9	4.74337861188046\\
%};
%\addplot[forget plot, color=white!15!black] table[row sep=crcr] {%
%0.506666666666667	0\\
%9.49333333333333	0\\
%};
\addplot[black,fill=mycolor6, draw=black] coordinates {
(1, 0.000838148743364158) (2,0.000168814558434414) (3,0.000474171381969933) (4,0.281077849289718) (5,1.09727991471399) (6,2.19060943132545) (7,3.02077227478425) (8,3.1770341724405) (9, 4.74337861188046)
  };
\addlegendentry{common rate}

\end{axis}

\end{tikzpicture}%

%% file: Results_K=2_K=8.tex
% This file was created by matlab2tikz.
%
%The latest updates can be retrieved from
%  http://www.mathworks.com/matlabcentral/fileexchange/22022-matlab2tikz-matlab2tikz
%where you can also make suggestions and rate matlab2tikz.
%
%\definecolor{mycolor1}{rgb}{0.49412,0.18431,0.55686}%
%\definecolor{mycolor2}{rgb}{0.07451,0.62353,1.00000}%
\definecolor{mycolor1}{rgb}{0.00000,0.44700,0.74100}%
\definecolor{mycolor3}{rgb}{0.00000,0.44706,0.74118}%
\definecolor{mycolor4}{rgb}{0.85098,0.32549,0.09804}%

\begin{tikzpicture}

\begin{axis}[%
width=4.521in,
height=3.566in,
at={(0.758in,0.481in)},
scale only axis,
xmin=0,
xmax=40,
xlabel={DL power [dB]},
ymin=0,
ymax=14,
ytick={0,2,4,6,8,10,12,14},
xticklabel style={font=\Large},
yticklabel style={font=\Large},
xlabel style={font=\color{white!15!black},font=\Large},
ylabel style={font=\color{white!15!black},font=\Large},
ylabel={achievable sum rate [bpcu]},
axis background/.style={fill=white},
xmajorgrids,
ymajorgrids,
legend style={at={(0.12,0.664)}, anchor=south west, legend cell align=left, align=left, draw=white!15!black,font=\Large}
]
\addplot [color=blue, line width=2pt,mark=star, mark options={solid, blue}, mark size=4pt]
  table[row sep=crcr]{%
-20	0.0118362829672138\\
-15	0.036877352015387\\
-10	0.156087763878904\\
-5	0.605226477122011\\
0	1.85031926811014\\
5	3.35416606341688\\
10	4.90967018739461\\
15	6.72629198734493\\
20	8.29035391249479\\
25	9.76477856768986\\
30	10.9932893517921\\
35	12.1907375209906\\
40	13.2926729097127\\
};
\addlegendentry{$K=2$, w/ RS}

\addplot [color=blue, dashed, line width=2pt,mark=star, mark options={solid, blue}, mark size=4pt]
  table[row sep=crcr]{%
-20	0.0118364868418082\\
-15	0.0378049062735164\\
-10	0.157607333509493\\
-5	0.60749538213023\\
0	1.85130609182191\\
5	3.3800377862695\\
10	4.88861421256252\\
15	6.34063672905989\\
20	7.10392983110652\\
25	7.56980837119454\\
30	7.72286409100566\\
35	7.78302521071429\\
40	7.8241179143012\\
};
\addlegendentry{$K=2$, w/o RS}

\addplot [color=red, line width=2pt,mark=triangle, mark options={solid, red}, mark size=4pt]
  table[row sep=crcr]{%
-20	0.0116473801034496\\
-15	0.0332632292082249\\
-10	0.108486543274632\\
-5	0.375037133925953\\
0	1.28765075142379\\
5	3.21547366941402\\
10	5.13767807840947\\
15	6.97052178048881\\
20	8.82214344234989\\
25	9.87984158907249\\
30	10.9190836996037\\
35	11.8157310326697\\
40	12.6133745878222\\
};
\addlegendentry{$K=8$, w/ RS}

\addplot [color=red, dashed, line width=2pt,mark=triangle, mark options={solid, red}, mark size=4pt]
  table[row sep=crcr]{%
-20	0.0116476417186512\\
-15	0.0332638365519673\\
-10	0.108487727441362\\
-5	0.375036658516005\\
0	1.28910795820109\\
5	3.21230653788758\\
10	5.13582555661092\\
15	6.97058866248124\\
20	8.82213940046\\
25	9.82571927738673\\
30	10.2455955778216\\
35	10.4092344692445\\
40	10.4730397038312\\
};
\addlegendentry{$K=8$, w/o RS}

\end{axis}

\end{tikzpicture}%

%% file: Results_M=32_M=16.tex
% This file was created by matlab2tikz.
%
%The latest updates can be retrieved from
%  http://www.mathworks.com/matlabcentral/fileexchange/22022-matlab2tikz-matlab2tikz
%where you can also make suggestions and rate matlab2tikz.
%
\definecolor{mycolor1}{rgb}{0.00000,0.44700,0.74100}%
\definecolor{mycolor2}{rgb}{0.92900,0.69400,0.12500}%
\definecolor{mycolor3}{rgb}{0.92941,0.69412,0.12549}%
\begin{tikzpicture}

\begin{axis}[%
width=4.521in,
height=3.566in,
at={(0.758in,0.481in)},
scale only axis,
xmin=0,
xmax=40,
xticklabel style={font=\Large},
yticklabel style={font=\Large},
xlabel style={font=\color{white!15!black},font=\Large},
ylabel style={font=\color{white!15!black},font=\Large},
xlabel={DL power [dB]},
ytick={0,2,4,6,8,10,12,14},
ymin=0,
ymax=14,
ylabel={achievable sum rate [bpcu]},
axis background/.style={fill=white},
xmajorgrids,
ymajorgrids,
legend style={at={(0.12,0.664)}, anchor=south west, legend cell align=left, align=left, draw=white!15!black,font=\Large}
]
\addplot [color=blue, line width=2pt, mark=triangle, mark options={solid, blue}, mark size=4pt]
  table[row sep=crcr]{%
0	1.61484102318866\\
5	3.17010165702113\\
10	4.47917489818705\\
15	6.52351661917866\\
20	7.96518213031595\\
25	9.25299668540663\\
30	10.3792663373252\\
35	11.3879746581026\\
40	12.1347722641421\\
};
\addlegendentry{M=32, w/ RS}

\addplot [color=blue, dashed, line width=2pt, mark=triangle, mark options={solid, blue}, mark size=4pt]
  table[row sep=crcr]{%
0	1.61615194044346\\
5	3.17008803460328\\
10	4.47921726159916\\
15	6.52354325051121\\
20	7.80760822912359\\
25	8.492839931875\\
30	8.71766650849418\\
35	8.80926413185988\\
40	8.83929813420348\\
};
\addlegendentry{M=32, w/o RS}

\addplot [color=red, line width=2pt,mark=star, mark options={solid, red}, mark size=4pt]
  table[row sep=crcr]{%
0	0.99019737247044\\
5	2.25039048907012\\
10	3.02925851003735\\
15	4.17436091315526\\
20	5.4135441304857\\
25	6.42582430004644\\
30	7.54521537190556\\
35	8.36928969702144\\
40	9.38031723228867\\
};
\addlegendentry{M=16, w/ RS}

\addplot [color=red, dashed, line width=2pt,mark=star, mark options={solid, red}, mark size=4pt]
  table[row sep=crcr]{%
0	0.991531307404306\\
5	2.25040140321041\\
10	3.02676386813875\\
15	4.09918288034455\\
20	4.96199940888763\\
25	5.39430260866415\\
30	5.58095421888137\\
35	5.64091368378805\\
40	5.65874323591817\\
};
\addlegendentry{M=16, w/o RS}

\end{axis}

\end{tikzpicture}%

%% file: Bilin_IWMMSE_RS_vs_noRS.tex
% This file was created by matlab2tikz.
%
%The latest updates can be retrieved from
%  http://www.mathworks.com/matlabcentral/fileexchange/22022-matlab2tikz-matlab2tikz
%where you can also make suggestions and rate matlab2tikz.
%
\definecolor{mycolor1}{rgb}{0.00000,0.44700,0.74100}%
\definecolor{mycolor2}{rgb}{1.0,0.07,0.65}%
\definecolor{mycolor3}{rgb}{0.92900,0.69400,0.12500}%
\definecolor{mycolor4}{rgb}{0.49400,0.18400,0.55600}%
\definecolor{mycolor5}{rgb}{0.36600,0.5400,0.10800}%
\definecolor{mycolor6}{rgb}{0.30100,0.74500,0.93300}%
\begin{tikzpicture}

\begin{axis}[%
width=4.521in,
height=3.566in,
at={(0.758in,0.481in)},
scale only axis,
xmajorgrids,
ymajorgrids,
xmin=0,
xmax=40,
ymin=2,
ymax=16,
xticklabel style={font=\Large},
yticklabel style={font=\Large},
xlabel style={font=\color{white!15!black},font=\Large},
ylabel style={font=\color{white!15!black},font=\Large},
xlabel={DL power [dB]},
ylabel={achievable sum rate [bpcu]},
axis background/.style={fill=white},
legend style={at={(0.9,0.3)},legend cell align=left, align=left, draw=white!15!black,font=\Large}
]
\addplot [color=blue, line width=2.0pt,mark=o, mark options={solid, blue}, mark size=4pt]
  table[row sep=crcr]{%
0	3.41316990712724\\
5	5.58830868705795\\
10	7.91493546148591\\
15	9.79351608509658\\
20	11.1303474838492\\
25	12.407690445371\\
30	13.589273452135\\
35	14.2038859855199\\
40	15.8623926005369\\
};
\addlegendentry{bilin. prec. w/ RS}

\addplot [color=blue, dashed, line width=2.0pt,mark=o, mark options={solid, blue}, mark size=4pt]
  table[row sep=crcr]{%
0	3.40792413844249\\
5	5.58899348947436\\
10	7.91514675823873\\
15	9.80426952886771\\
20	10.8414409091281\\
25	11.1496506081529\\
30	11.2013829813646\\
35	11.2752788288583\\
40	11.462833952141\\
};
\addlegendentry{bilin. prec. w/o RS}

\addplot [color=mycolor3, line width=2pt,mark=square, mark options={solid, mycolor3}, mark size=4pt]
  table[row sep=crcr]{%
0	2.94696992878553\\
5	4.69056855254683\\
10	6.28283295917271\\
15	7.93855159674063\\
20	9.11440993750118\\
25	9.93113893517571\\
30	11.3145038456826\\
35	12.2974441699474\\
40	13.6461180452503\\
};
\addlegendentry{IWMMSE prec. w/ RS}

\addplot [color=mycolor3, dashed, line width=2.0pt,mark=square, mark options={solid, mycolor3}, mark size=4pt]
  table[row sep=crcr]{%
0	2.52521229876802\\
5	4.13579833797946\\
10	5.68351343108407\\
15	6.75771131348445\\
20	7.45873588012753\\
25	7.43483540045436\\
30	7.4024801362582\\
35	7.57366604634819\\
40	7.60057715990504\\
};
\addlegendentry{IWMMSE prec. w/o RS}

\end{axis}

\end{tikzpicture}%

%% file: Bilin_IWMMSE_Known_vs_Estimated_Cov_Matrices.tex
% This file was created by matlab2tikz.
%
%The latest updates can be retrieved from
%  http://www.mathworks.com/matlabcentral/fileexchange/22022-matlab2tikz-matlab2tikz
%where you can also make suggestions and rate matlab2tikz.
%
\definecolor{mycolor1}{rgb}{0.00000,0.44700,0.74100}%
\definecolor{mycolor2}{rgb}{1.0,0.07,0.65}%
\definecolor{mycolor3}{rgb}{0.92900,0.69400,0.12500}%

\begin{tikzpicture}

\begin{axis}[%
width=4.521in,
height=3.566in,
at={(0.758in,0.481in)},
scale only axis,
xmajorgrids,
ymajorgrids,
xmin=0,
xmax=40,
ymin=2,
ymax=16,
xticklabel style={font=\Large},
yticklabel style={font=\Large},
xlabel style={font=\color{white!15!black},font=\Large},
ylabel style={font=\color{white!15!black},font=\Large},
xlabel={DL power [dB]},
ylabel={achievable sum rate [bpcu]},
axis background/.style={fill=white},
legend style={at={(0.95,0.3)},legend cell align=left, align=left, draw=white!15!black,font=\large}
]
\addplot [color=mycolor1, line width=2pt,,mark=o, mark options={solid, blue}, mark size=4pt]
  table[row sep=crcr]{%
0	3.41316990712724\\
5	5.58830868705795\\
10	7.91493546148591\\
15	9.79351608509658\\
20	11.1303474838492\\
25	12.407690445371\\
30	13.589273452135\\
35	14.2038859855199\\
40	15.8623926005369\\
};
\addlegendentry{bilin. prec. known cov.}

\addplot [color=mycolor1, dashdotted, line width=2.0pt,mark=o, mark options={solid, blue}, mark size=4pt]
  table[row sep=crcr]{%
0	3.03785631100595\\
5	4.663084011521\\
10	6.46762443427642\\
15	7.83661597766616\\
20	9.20500104918324\\
25	10.4207635808869\\
30	11.3135496926201\\
35	11.7617081089682\\
40	13.5950114744666\\
};
\addlegendentry{bilin. prec. est. cov.}

\addplot [color=mycolor3, line width=2.0pt,mark=square, mark options={solid, mycolor3}, mark size=4pt]
  table[row sep=crcr]{%
0	2.94696992878553\\
5	4.69056855254683\\
10	6.28283295917271\\
15	7.93855159674063\\
20	9.11440993750118\\
25	9.93113893517571\\
30	11.3145038456826\\
35	12.2974441699474\\
40	13.6461180452503\\
};
\addlegendentry{IWMMSE prec. known cov.}

\addplot [color=mycolor3,dashdotted, line width=2pt,mark=square, mark options={solid, mycolor3}, mark size=4pt]
  table[row sep=crcr]{%
0	2.58203774999649\\
5	3.79993511319175\\
10	4.98866655129418\\
15	6.20515122730773\\
20	7.25423902209803\\
25	8.43785670024313\\
30	9.66254162951656\\
35	10.616949481918\\
40	12.1552882908893\\
};
\addlegendentry{IWMMSE prec. est. cov.}

\end{axis}

\end{tikzpicture}%

%% file: Appendix1.tex
We prove the results presented in \eqref{eq:SINRc} and \eqref{eq:SINRp}.
We begin with deriving the closed-form expression for the private $\gamma_{\text{p},k}$. Recall the channel estimate $\boldsymbol{y}_k=\boldsymbol{\Phi}^\HH \boldsymbol{h}_k + \boldsymbol{n}_k$ [see \eqref{eq:yk}], where the noise term $\boldsymbol{n}_k \sim \mathcal{N}_\mathbb{C}(\boldsymbol{0},\sigma_n^2\boldsymbol{I})$ and the $\boldsymbol{h}_k$ are mutually independent. Furthermore, it holds that the private precoder of user $k$ is given by $\boldsymbol{p}_{\text{p},k}=\boldsymbol{A}_{\text{p},k}\boldsymbol{y}_k$.\\
The numerator of the private SINR can be expressed as, where $\boldsymbol{C}_k$ is the covariance matrix of the channel vector $\boldsymbol{h}_k$
\begin{align}
	|\mathbb{E}[\boldsymbol{h}_k^\HH \boldsymbol{p}_{\text{p},k}]|^2&=|\mathbb{E}[\boldsymbol{h}_k^\HH \boldsymbol{A}_{\text{p},k}\boldsymbol{y}_k]|^2\nonumber\\
	&=|\mathbb{E}[\boldsymbol{h}_k^\HH \boldsymbol{A}_{\text{p},k}(\boldsymbol{\Phi}^\HH \boldsymbol{h}_k + \boldsymbol{n}_k)]|^2\nonumber\\
	&= |\text{tr}(\boldsymbol{A}_{\text{p},k}\boldsymbol{\Phi}^\HH\boldsymbol{C}_k)|^2 \label{eq:1term}
\end{align}
due to the independence of $\boldsymbol{h}_k$ and $\boldsymbol{n}_k$.\\
Since it holds that $\text{var}(\boldsymbol{h}_k^\HH \boldsymbol{p}_{\text{p},k})=\mathbb{E}[|\boldsymbol{h}_k^\HH \boldsymbol{p}_{\text{p},k}|^2]-|\mathbb{E}[\boldsymbol{h}_k^\HH \boldsymbol{p}_{\text{p},k}]|^2$, we only need to concentrate on the first term of this expression since the second term can be found in \eqref{eq:1term}:
\begin{align}
\mathbb{E}[|\boldsymbol{h}_k^\HH \boldsymbol{p}_{\text{p},k}|^2]&=\mathbb{E}[|\boldsymbol{h}_k^\HH \boldsymbol{A}_{\text{p},k}\boldsymbol{y}_k|^2] \nonumber\\
&=\mathbb{E}[|\boldsymbol{h}_k^\HH\boldsymbol{A}_{\text{p},k}\boldsymbol{\Phi}^\HH\boldsymbol{h}_k|^2] + \sigma_n^2\text{tr}(\boldsymbol{A}_{\text{p},k}^\HH \boldsymbol{C}_k\boldsymbol{A}_{\text{p},k}). \label{eq:Lemma3}
\end{align}

To compute the first term of \eqref{eq:Lemma3}, we use \cite[\textit{Lemma 2}]{Lemma3} and obtain, therefore,
\begin{align}
\mathbb{E}[|\boldsymbol{h}_k^\HH \boldsymbol{p}_{\text{p},k}|^2]&=|\text{tr}(\boldsymbol{A}_{\text{p},k}\boldsymbol{\Phi}^\HH\boldsymbol{C}_k)|^2+\text{tr}(\boldsymbol{A}_{\text{p},k}\boldsymbol{\Phi}^\HH\boldsymbol{C}_k\boldsymbol{\Phi}\boldsymbol{A}_{\text{p},k}^\HH\boldsymbol{C}_k)\nonumber \\
&+\sigma_n^2\text{tr}(\boldsymbol{A}_{\text{p},k}^\HH \boldsymbol{C}_k\boldsymbol{A}_{\text{p},k})\\
&=|\text{tr}(\boldsymbol{A}_{\text{p},k}\boldsymbol{\Phi}^\HH\boldsymbol{C}_k)|^2+\text{tr}(\boldsymbol{A}_{\text{p},k}\boldsymbol{C}_{\boldsymbol{y}_k} \boldsymbol{A}_{\text{p},k}^\HH\boldsymbol{C}_k)
\end{align}

Next, we consider the term $\mathbb{E}[|\boldsymbol{h}_k^\HH \boldsymbol{p}_{\text{p},i}|^2]$ for $i\neq k$. Following similar steps, we obtain
\begin{align}
\mathbb{E}[|\boldsymbol{h}_k^\HH \boldsymbol{p}_{\text{p},i}|^2]
=\text{tr}(\boldsymbol{A}_{\text{p},i}\boldsymbol{C}_{\boldsymbol{y}_i} \boldsymbol{A}_{\text{p},i}^\HH\boldsymbol{C}_k)
\end{align}
where we used the independence of $\boldsymbol{h}_k$ and $\boldsymbol{h}_i$ for $k\neq i$.\\
With all these results, the private SINR of user $k$ can be written as
\begin{equation}
    \gamma_{\text{p},k}=\frac{|\text{tr}(\boldsymbol{A}_{\text{p},k}\boldsymbol{\Phi}^\HH\boldsymbol{C}_k)|^2}{\sum_{i=1}^K \text{tr}(\boldsymbol{A}_{\text{p},i}\boldsymbol{C}_{\boldsymbol{y}_i} \boldsymbol{A}_{\text{p},i}^\HH\boldsymbol{C}_k)+1}.
\end{equation}
If we now consider the vectorized forms of the transformation matrices
\begin{equation}
\label{eq:apk}
    \boldsymbol{a}_{\text{p},k}=\text{vec}(\boldsymbol{A}_{\text{p},k})
\end{equation}
and the covariance matrices $\boldsymbol{c}_k=\text{vec}(\boldsymbol{C}_k)$
by using the following identities \cite[T3.9, T3.4, T2.13]{Brewer}
\begin{equation}
\label{eq:TrVec}
  \begin{aligned}
    \text{tr}(\boldsymbol{A}^\HH \boldsymbol{B})&=\text{vec}^\HH(\boldsymbol{A}) \text{vec}(\boldsymbol{B})\\ \text{vec}(\boldsymbol{C}\boldsymbol{A}\boldsymbol{D})&= (\boldsymbol{D}^\TT\otimes \boldsymbol{C})\: \text{vec}(\boldsymbol{A})
\end{aligned}  
\end{equation}
the private SINR of user $k$ can be formulated as
\begin{align}
    \gamma_{\text{p},k}&=\frac{|\boldsymbol{c}_k^\HH (\boldsymbol{\Phi}^*\otimes \boldsymbol{I}) \boldsymbol{a}_{\text{p},k}|^2}{\sum_{i=1}^K \boldsymbol{a}_{\text{p},i}^\HH (\boldsymbol{C}_{\boldsymbol{y}_i}^\TT \otimes  \boldsymbol{C}_k )\boldsymbol{a}_{\text{p},i} +1 }\nonumber \\
    &=\frac{|\boldsymbol{q}_k^\HH  \boldsymbol{a}_{\text{p},k}|^2}{\sum_{i=1}^K \boldsymbol{a}_{\text{p},i}^\HH \boldsymbol{Q}_{i,k} \boldsymbol{a}_{\text{p},i} +1 }
    \label{eq:SINRp2}
\end{align}
where we defined the vector $\boldsymbol{q}_k=(\boldsymbol{\Phi}^\TT\otimes \boldsymbol{I}_M) \boldsymbol{c}_k$ and the matrix $\boldsymbol{Q}_{i,k}=\boldsymbol{C}_{\boldsymbol{y}_i}^\TT \otimes  \boldsymbol{C}_k$. 

In order to derive a compact closed-form expression for the common SINR of user $k$, we first define the vectors $\boldsymbol{h}$ and $\boldsymbol{n}$, where we stack the channel vectors and noise vectors of all users, respectively, such that 
\begin{equation}
    \begin{aligned}
    &\boldsymbol{h}=[\boldsymbol{h}_1^\TT,\dots, \boldsymbol{h}_K^\TT]^\TT, &&\quad \boldsymbol{n}=[\boldsymbol{n}_1^\TT,\dots, \boldsymbol{n}_K^\TT]^\TT\\
    &\boldsymbol{h}_k=\boldsymbol{E}_k\boldsymbol{h}, &&\quad \boldsymbol{n}_k=\boldsymbol{E}_k\boldsymbol{n}
    \end{aligned}
\end{equation}
where $\boldsymbol{E}_k=\boldsymbol{e}_k^\TT\otimes \boldsymbol{I}_M$ is an $M$ times $KM$ selection matrix.\\
The vector $\boldsymbol{y}$ of all users' channel observations can be therefore written as
\begin{equation}
    \boldsymbol{y}=\underbrace{(\boldsymbol{I}_K \otimes \boldsymbol{\Phi}^\HH)}_{\boldsymbol{D}} \boldsymbol{h} + \boldsymbol{n}
\end{equation}
whose covariance matrix is given by
\begin{equation}
    \boldsymbol{C}_{\boldsymbol{y}}=\boldsymbol{D} \boldsymbol{C}_{\boldsymbol{h}}  \boldsymbol{D}^\HH + \boldsymbol{C}_{\boldsymbol{n}}
\end{equation}
with $\boldsymbol{C}_{\boldsymbol{h}}=\mathbb{E}[\boldsymbol{h}\boldsymbol{h}^\HH ]$ and $\boldsymbol{C}_{\boldsymbol{n}}=\mathbb{E}[\boldsymbol{n}\boldsymbol{n}^\HH ]$ being the block diagonal covariance matrices of $\boldsymbol{h}$ and $\boldsymbol{n}$, respectively.\\
Following similar steps as done for the derivation of the private SINR, we obtain the following terms for the common SINR
\begin{equation}
    \mathbb{E}[\boldsymbol{h}_k^\HH \boldsymbol{p}_\text{c}] =\text{tr}(\boldsymbol{C}_{\boldsymbol{h}}\boldsymbol{E}_k^\TT \boldsymbol{A}_\text{c} \boldsymbol{D})
    =\boldsymbol{c}_{\boldsymbol{h}}^\HH (\boldsymbol{D}^\TT\otimes \boldsymbol{E}_k^\TT) \boldsymbol{a}_\text{c}
    =\boldsymbol{z}_k^\HH\boldsymbol{a}_\text{c}
\end{equation}
where we used the vectorized forms $\boldsymbol{c}_{\boldsymbol{h}}=\text{vec}(\boldsymbol{C}_{\boldsymbol{h}})$ and $\boldsymbol{a}_\text{c}=\text{vec}(\boldsymbol{A}_\text{c})$, and defined the vector $\boldsymbol{z}_k=(\boldsymbol{D}^*\otimes \boldsymbol{E}_k) \boldsymbol{c}_{\boldsymbol{h}}$.\\
For $\text{var}(\boldsymbol{h}_k^\HH\boldsymbol{p}_\text{c})$, we need
\begin{align}
    \mathbb{E}[|\boldsymbol{h}_k^\HH \boldsymbol{p}_\text{c}|^2]&=\text{tr}(\boldsymbol{C}_{\boldsymbol{h}}\boldsymbol{E}_k^\TT \boldsymbol{A}_\text{c} \boldsymbol{D})\nonumber\\
    &+\text{tr}(\underbrace{\boldsymbol{E}_k\boldsymbol{C}_{\boldsymbol{h}}\boldsymbol{E}_k^\HH}_{\boldsymbol{C}_k} \boldsymbol{A}_\text{c} (\boldsymbol{D}\boldsymbol{C}_{\boldsymbol{h}} \boldsymbol{D}^\HH+\boldsymbol{C}_{\boldsymbol{n}})\boldsymbol{A}_\text{c}^\HH)\\
    &=|\boldsymbol{z}_k^\HH \boldsymbol{a}_\text{c}|^2+\boldsymbol{a}_\text{c}^\HH \underbrace{(\boldsymbol{C}_{\boldsymbol{y}}^\TT\otimes \boldsymbol{C}_k)}_{\boldsymbol{Z}_k}\boldsymbol{a}_\text{c}.\label{eq:Zk}
\end{align} 
%where $\boldsymbol{z}_k=(\boldsymbol{D}^*\otimes \boldsymbol{E}_k)\boldsymbol{c}_h$.\\
With the previous result for the denominator in \eqref{eq:SINRp2}, the common SINR of user $k$ can be written in its compact form as
\begin{equation}
    \gamma_{\text{c},k}=\frac{|\boldsymbol{z}_k^\HH\boldsymbol{a}_\text{c}|^2}{\boldsymbol{a}_\text{c}^\HH\boldsymbol{Z}_k\boldsymbol{a}_\text{c}+ |\boldsymbol{q}_k^\HH  \boldsymbol{a}_{\text{p},k}|^2+ \sum_{i=1}^K \boldsymbol{a}_{\text{p},i}^\HH \boldsymbol{Q}_{i,k}\boldsymbol{a}_{\text{p},i} +1}.
\end{equation}

%% file: Appendix2.tex
 For fixed step size $u$, the parameters $v$ and $\bs{a}_\tc^{(n),\perp}$ in \eqref{eq:AcUpdate} are chosen such that: \\
\textbf{1.} The power constraint $\bs{a}_\tc^\HH \bs{F} \bs{a}_\tc \leq \alpha_\tc P_\text{dl}$ is satisfied with equality at iteration $n+1$ for $\bs{a}_\tc=\bs{a}_\tc^{(n+1)}$.
\begin{align*}
	&\bs{a}_\tc^{(n+1),\HH}\bs{F} \bs{a}_\tc^{(n+1)}		\\
	&=(1-u)^2\alpha_\tc P_\text{dl}+ |v|^2 \|\bs{F}^{\frac{1}{2}}\bs{Z}_{\ell(n)}^{-\frac{1}{2}}\bs{a}_\tc^{(n),\perp}\|_2^2\\
	&+2(1-\alpha)\Re\{\beta \bs{a}_\tc^{(n),\HH} \bs{F} \bs{Z}_{\ell(n)}^{-\frac{1}{2}} \bs{a}_\tc^{(n),\perp} \}	\overset{!}{=} \alpha_\tc P_\text{dl} \label{eq:Condition}
	\end{align*} 
	If we set $\|\bs{F}^{\frac{1}{2}}\bs{Z}_{\ell(n)}^{-\frac{1}{2}}\bs{a}_\tc^{(n),\perp}\|_2^2$ to $1$, which corresponds to a normalization of the vector $\bs{a}_\tc^{(n),\perp}$, and $\Re\{\beta \bs{a}_\tc^{(n),\HH} \bs{F} \bs{Z}_{\ell(n)}^{-\frac{1}{2}} \bs{a}_\tc^{(n),\perp} \} $ to zero, by imposing the orthogonality constraint 
	\begin{equation}
	    \bs{a}_\tc^{(n),\perp} \perp \bs{Z}_{\ell(n)}^{-\frac{1}{2},\HH }  \bs{F}^\HH \bs{a}_\tc^{(n)}
	\end{equation}
	the magnitude of $v$ is given by
	\begin{equation}
	    |v|=\sqrt{\alpha_\tc P_\text{dl}(2u-u^2)}.
	    \label{eq:MagV}
	\end{equation}
	since $\bs{a}_\tc^{(n),\HH}\bs{F} \bs{a}_\tc^{(n)}=\alpha_\tc P_\text{dl}$.\\
\textbf{2.} The left-hand-side of the SINR constraint in \eqref{eq:ComOpt3} is maximized. When inserting the update in \eqref{eq:AcUpdate}, the latter is given by
\begin{align}
    &2\Re\{\eta_{\ell(n)}^*\bs{z}_{\ell(n)}^\HH\bs{a}_\tc^{(n+1)} \} \nonumber\\
    &-|\eta_{\ell(n)}|^2\left(\bs{a}_\tc^{(n+1),\HH}\bs{Z}_{\ell(n)}\bs{a}_\tc^{(n+1)}+\sigma_{\ell(n)}^2\right)\\
    =&2(1-u)\Re\{\eta_{\ell(n)}^*\bs{z}_{\ell(n)}^\HH\bs{a}_\tc^{(n)} \}\label{eq:1stLine}\\
    -&|\eta_{\ell(n)}|^2\left((1-u)^2\bs{a}_\tc^{(n),\HH}\bs{Z}_{\ell(n)}\bs{a}_\tc^{(n)}+\sigma_{\ell(n)}^2\right)\label{eq:2ndLine}\\
    -& |\eta_{\ell(n)}|^2 |v|^2 \bs{a}_\tc^{(n),\perp, \HH}  \bs{Z}_{\ell(n)}^{-\frac{1}{2},\HH }\bs{Z}_{\ell(n)} \bs{Z}_{\ell(n)}^{-\frac{1}{2}} \bs{a}_\tc^{(n),\perp} \label{eq:3rdLine}\\ 
    +&2 |v|   \Re\Big\{\mathrm{e}^{\mathrm{j}\angle v} \bs{t}_\tc^{(n),\HH} \bs{a}_\tc^{(n),\perp}\Big\}\label{eq:4thLine}
\end{align} 
where we used for the last term \eqref{eq:4thLine} that $v=|v| \mathrm{e}^{\mathrm{j}\angle v}$. Furthermore, we defined the vector $\bs{t}_\tc^{(n)}$ such that
\begin{equation}
    \bs{t}_\tc^{(n)}=\eta_{\ell(n)}\bs{Z}_{\ell(n)}^{-\frac{1}{2},\HH}\bs{z}_{\ell(n)}-|\eta_{\ell(n)}|^2(1-u) \bs{Z}_{\ell(n)}^{\frac{1}{2}}\bs{a}_\tc^{(n)}
\end{equation}
The terms in \eqref{eq:1stLine} and \eqref{eq:2ndLine} depend only on $\bs{a}_\tc^{(n)}$, the common transformation vector from iteration $n$, and are therefore not taken into account when maximizing the whole expression.\\
Due to the transformation applied to $\bs{a}_\tc^{(n),\perp}$ via the matrix $\bs{Z}_{\ell(n)}^{-\frac{1}{2}}$, we can see that the third term in \eqref{eq:3rdLine} simplifies to $ |\eta_{\ell(n)}|^2 |v|^2 \|\bs{a}_\tc^{(n),\perp}\|^2$. Since the norm of $\bs{a}_\tc^{(n),\perp}$ is already fixed by the transmit power constraint, the only term we have to maximize is the one given by \eqref{eq:4thLine}. Here, the maximization is done with respect to $\bs{a}_\tc^{(n),\perp}$ and $\angle v$, the phase of $v$ [the magnitude is already fixed due to the power constraint, cf. \eqref{eq:MagV}]. \\
For fixed $\bs{a}_\tc^{(n),\perp}$, $\angle v$ is chosen such that the real part in \eqref{eq:4thLine} is maximized and thus it is given by
\begin{equation}
     \angle v=-\angle \bs{t}_\tc^{(n),\HH}\bs{a}_\tc^{(n),\perp}.
\end{equation}
With this choice for $\angle v$, the real part in \eqref{eq:4thLine} can be written as
\begin{equation}
    |\bs{t}_\tc^{(n),\HH}\bs{a}_\tc^{(n),\perp}|
\end{equation}
which is maximized by choosing $\bs{a}_\tc^{(n),\perp}$ as "co-linear" as possible to $\bs{t}_\tc^{(n)}$.